\documentstyle[psfig]{mn}

\newif\ifAMStwofonts

\def\kms{\relax \ifmmode {\,\rm km\,s}^{-1}\else \,km\,s$^{-1}$\fi}
\def\ha{\relax \ifmmode {\rm H}\alpha\else H$\alpha$\fi}
\def\hb{\relax \ifmmode {\rm H}\beta\else H$\beta$\fi}
\def\hi{\relax \ifmmode {\rm H\,{\sc i}}\else H\,{\sc i}\fi}
\def\hii{\relax \ifmmode {\rm H\,{\sc ii}}\else H\,{\sc ii}\fi}
\def\h2{\relax \ifmmode {\rm H}_2\else H$_2$\fi}
\def\lha{\relax \ifmmode L_{{\rm H}\alpha}\else $L_{{\rm H}\alpha}$\fi}
\def\shi{\relax \ifmmode \sigma_{{\rm HI}}\else $\sigma_{\rm HI}$\fi}
\def\sh2{\relax \ifmmode \sigma_{{\rm H}_2}\else $\sigma_{{\rm H}_2}$\fi}
\def\degr{\hbox{$^\circ$}}
\def\arcmin{\hbox{$^\prime$}}
\def\arcsec{\hbox{$^{\prime\prime}$}}
\def\deg{\hbox{$^\circ$}}

\def\fdg{\hbox{$.\!\!^\circ$}}
\def\fs{\hbox{$.\!\!^{\rm s}$}}
\def\farcm{\hbox{$.\mkern-4mu^\prime$}}
\def\farcs{\hbox{$.\!\!^{\prime\prime}$}}
\def\degd#1.#2{ #1\fdg#2 }                 
\def\mind#1.#2{ #1\farcm#2 }               
\def\secd#1.#2{ #1\farcs#2 }               
\def\hhh{\ifmmode {\rm ^h}              
         \else {${\rm ^h}$}
         \fi}
\def\sss{\ifmmode {\rm ^s}              
         \else {${\rm ^s}$}
         \fi}
\def\hms#1h#2m#3s{                      
                  \relax
                  \ifmmode #1^{\rm h}\,#2^{\rm m}\,#3^{\rm s}
                  \else \hbox{$#1^{\rm h}\,#2^{\rm m}\,#3^{\rm s}$}
                  \fi
                 }
\def\dms#1d#2m#3s{                      
                  \relax
                  #1\degr\,#2\arcmin\,#3\arcsec 
                 }
\def\hmsd#1h#2m#3.#4s{                  
                      \relax
                      \ifmmode #1^{\rm h}\,#2^{\rm m}\,#3\fs#4
                      \else \hbox{$#1^{\rm h}\,#2^{\rm m}\,#3\fs#4$}
                      \fi
                     }
\def\dmsd#1d#2m#3.#4s{                  
                      \relax
                      #1\degr\,#2\arcmin\,#3\farcs#4
                     }
\def\mag{\relax                          
        \ifmmode ^{\rm m}
        \else $^{\rm m}$
        \fi
       }
\def\magd#1.#2{                          
              \relax
              \ifmmode #1^{\rm m}
                       \hskip-0.55em.\hskip0.22em#2
              \else \hbox{#1$^{\rm m}
                    \hskip-0.55em.\hskip0.22em$#2}
              \fi
             }

\ifoldfss
  \ifCUPmtlplainloaded \else
    \NewTextAlphabet{textbfit} {cmbxti10} {}
    \NewTextAlphabet{textbfss} {cmssbx10} {}
    \NewMathAlphabet{mathbfit} {cmbxti10} {} 
    \NewMathAlphabet{mathbfss} {cmssbx10} {} 
  \fi
  \ifAMStwofonts
    \ifCUPmtlplainloaded \else
      \NewSymbolFont{upmath} {eurm10}
      \NewSymbolFont{AMSa} {msam10}
      \NewMathSymbol{\upi}     {0}{upmath}{19}
      \NewMathSymbol{\umu}     {0}{upmath}{16}
      \NewMathSymbol{\upartial}{0}{upmath}{40}
      \NewMathSymbol{\leqslant}{3}{AMSa}{36}
      \NewMathSymbol{\geqslant}{3}{AMSa}{3E}

      \let\geq=\geqslant 
    \fi
  \fi
\fi 

\ifnfssone
  \newmathalphabet{\mathit}
  \addtoversion{normal}{\mathit}{cmr}{m}{it}
  \addtoversion{bold}{\mathit}{cmr}{bx}{it}
  \newmathalphabet{\mathbfit} 
  \addtoversion{normal}{\mathbfit}{cmr}{bx}{it}
  \addtoversion{bold}{\mathbfit}{cmr}{bx}{it}
  \newmathalphabet{\mathbfss} 
  \addtoversion{normal}{\mathbfss}{cmss}{bx}{n}
  \addtoversion{bold}{\mathbfss}{cmss}{bx}{n}
  \ifAMStwofonts
    \ifCUPmtlplainloaded \else
      \UseAMStwoboldmath
      \makeatletter
      \new@mathgroup\upmath@group
      \define@mathgroup\mv@normal\upmath@group{eur}{m}{n}
      \define@mathgroup\mv@bold\upmath@group{eur}{b}{n}
      \edef\UPM{\hexnumber\upmath@group}
      \new@mathgroup\amsa@group
      \define@mathgroup\mv@normal\amsa@group{msa}{m}{n}
      \define@mathgroup\mv@bold\amsa@group{msa}{m}{n}
      \edef\AMSa{\hexnumber\amsa@group}
      \makeatother
      \mathchardef\upi="0\UPM19
      \mathchardef\umu="0\UPM16
      \mathchardef\upartial="0\UPM40
      \mathchardef\leqslant="3\AMSa36
      \mathchardef\geqslant="3\AMSa3E

      \let\geq=\geqslant 
    \fi
  \fi
\fi 

\ifnfsstwo
  \DeclareMathAlphabet{\mathbfit}{OT1}{cmr}{bx}{it}
  \SetMathAlphabet\mathbfit{bold}{OT1}{cmr}{bx}{it}
  \DeclareMathAlphabet{\mathbfss}{OT1}{cmss}{bx}{n}
  \SetMathAlphabet\mathbfss{bold}{OT1}{cmss}{bx}{n}
  \ifAMStwofonts
    \ifCUPmtlplainloaded \else
      \DeclareSymbolFont{UPM}{U}{eur}{m}{n}
      \SetSymbolFont{UPM}{bold}{U}{eur}{b}{n}
      \DeclareSymbolFont{AMSa}{U}{msa}{m}{n}
      \DeclareMathSymbol{\upi}{0}{UPM}{"19}
      \DeclareMathSymbol{\umu}{0}{UPM}{"16}
      \DeclareMathSymbol{\upartial}{0}{UPM}{"40}
      \DeclareMathSymbol{\leqslant}{3}{AMSa}{"36}
      \DeclareMathSymbol{\geqslant}{3}{AMSa}{"3E}

      \let\geq=\geqslant 
    \fi
  \fi
\fi 

\ifCUPmtlplainloaded \else
  \ifAMStwofonts \else 
    \def\upi{\pi}
    \def\umu{\mu}
    \def\upartial{\partial}
  \fi
\fi

\title[Circumnuclear regions in barred spiral galaxies
II.]{Circumnuclear regions in barred spiral galaxies II. Relations to
host galaxies} 
\author[J.~H. Knapen et al.]{J. H.  Knapen$^{1,2}$, D. P\'erez-Ram\'\i
rez$^{3}$, S. Laine $^{4}$\\
$^1$ Isaac Newton Group of Telescopes, Apartado 321, E-38700 Santa Cruz
de La Palma, Spain\\
$^2$ Department of Physical Sciences, University of Hertfordshire,
Hatfield, Herts AL10 9AB, UK\\
$^3$ Department of Physics and Astronomy, Michigan Technological
University, 1400 Townsend Dr., Houghton, MI 49931, USA\\
$^4$ Space Telescope Science Institute, 3700 San Martin Drive,
Baltimore, MD 21218, USA}

\date{Accepted July 5, 2002;
      Received;
      in original form}

\pagerange{\pageref{firstpage}--\pageref{lastpage}}
\pubyear{2000}

\begin{document}

\maketitle

\label{firstpage}

\begin{abstract}

We present optical broad- and narrow-band imaging of a sample of a dozen
barred galaxies. These images are analyzed in conjunction with our
previously published near-infrared imaging of their central regions and
with literature values for, e.g., bar strengths and the total star
formation activity of the galaxies. We present $B$, $I$ and H$\alpha$
images, and radial profiles derived from these, to infer geometric and
dynamical parameters of the structural components of the galaxies, such
as bar lengths, bar ellipticities, and location of star formation and
dust. We find that the more centrally concentrated the \ha\ emission in
a galaxy is, i.e., the higher the fraction of star formation originating
in the circumnuclear region, the higher the overall star formation rate,
as measured from far-infrared flux ratios.  Stronger bars host smaller
nuclear rings, but the strength of the bar does not correlate with
either the intrinsic ellipticity of the ring or the offset between the
position angles of the bar and the ring. We interpret these results in
comparison with modelling of gas inflow in the circumnuclear region, and
show that they were theoretically expected.  We confirm observationally,
and for the first time, the anti-correlation predicted from theory and
modelling between the degree of curvature of the bar dust lanes and the
strength of the bar, where stronger bars have straighter dust lanes.

\end{abstract}


\section{Introduction} 

Most galaxies are barred (e.g., Sellwood \& Wilkinson 1993; Knapen,
Shlosman \& Peletier 2000a) and a substantial fraction of barred
galaxies show enhanced star formation (SF) activity in or near their
centres, often in the form of complete or incomplete nuclear rings
(e.g., Buta \& Combes 1996; Knapen 1999). It is believed that there is a
causal connection between the existence of a bar and circumnuclear SF
activity. Through its non-axisymmetric potential, a bar can facilitate
gas inflow by extracting angular momentum from the gas through
gravitational torques. The inflowing gas may then accumulate in the
vicinity of inner Lindblad resonances (ILRs), triggering massive SF
(see, e.g., review by Shlosman 1999). It is therefore natural to infer
that the properties of the circumnuclear starburst region are connected
to those of the large-scale structure, most obviously the bar, and it is
this inference that we wish to test observationally in the present
paper.

Correlations have been established between morphological properties of
the stellar bar, such as size, shape and surface brightness (Martin
1995; Elmegreen \& Elmegreen 1985), the location of star forming regions
(Sersic \& Pastoriza 1967; Phillips 1993, 1996), the presence of nuclear
activity (Chapelon, Contini \& Davoust 1999; Knapen et al. 2000a;
Shlosman, Peletier \& Knapen 2000; Laine et al. 2002) and the Hubble
type of their host galaxy. Bars are found to be longer in early-type
barred galaxies (Martin 1995), longer bars are more elliptical (Martinet
\& Friedli 1997), bar radial profiles appear to be flatter in early-type
galaxies and exponential in late-type galaxies (Elmegreen
\& Elmegreen 1985), and there is some evidence that the more
elliptical the bar is, the higher is its overall SF rate (Aguerri
1999). There also appears to be some correlation between the location
of SF in galaxies and their morphological classification. Phillips
(1993) noted that early-type barred galaxies exhibit SF in rings but
neither in the bar nor the centre, whereas late-type barred spirals
display many star forming regions along their bars.

Sersic \& Pastoriza (1967) realised that enhanced SF is found in the
inner parts of some spiral galaxies, and that this SF is frequently
arranged into a ring or pseudo-ring pattern. Combes \& Gerin (1985) and
Athanassoula (1992a) found that such spiral galaxies are usually of
early type and host a bar. Rings in galaxies, including nuclear rings,
were extensively reviewed by Buta \& Combes (1996). Buta \& Crocker
(1993) compiled a catalog of rings in galaxies, including ten nuclear
rings studied in the present paper, and derived metric characteristics
for the rings. Case studies combining optical and near-infrared (NIR)
imaging with velocity fields and dynamical modelling show in detail how
the bar and the circumnuclear regions (CNRs) are part of the same
dynamical system. Gas is driven inward under the influence of a large
stellar bar, where it piles up near or between the ILRs, and possibly
results in massive SF. Examples of such case studies are those by Knapen
et al.  (1995a,b, 2000b) on M100 (=NGC~4321), Regan et al. (1996, 1997)
on NGC~1530, Buta, Crocker \& Byrd (1999) on ESO~565-11 (see also
Rautiainen \& Salo 2000), and Laine et al. (1999, 2001) and Jogee et
al. (2002a,b) on NGC~5248. In thin bars (high axial ratio) the gas
inflow may be most effective, but according to the simulations by Piner,
Stone \& Teuben (1995), nuclear rings will not form in such bars and gas
will flow directly to the nucleus.  Nuclear bars, spirals, or rings, are
frequently observed on scales of a kiloparsec or less, and can be
modelled as directly resulting from the dynamics of the overall system
(see Knapen et al. 2001 for many examples).

We have performed an imaging study of a dozen barred spirals with
evidence for the presence of star-forming CNRs, to investigate the
coupling between the CNRs and their host galaxies in a more statistical
manner.  Our main goal is to test the hypothesis (Knapen et al. 1995b;
P\'erez-Ram\'\i rez \& Knapen 1998) that the structure and dynamics of
the large-scale bar and disk of the host galaxies is intimately
connected with that of the inner, active region (either AGN or SF), and
thus determines whether such activity occurs in a given galaxy, and in
what form.  In Paper~I (P\'erez-Ram\'\i rez et al. 2000), we presented
NIR images of the central regions of a sample of 12 barred galaxies to
study the properties of circumnuclear structures in stars and dust that
exist at small scales (a few hundred pc to a few kpc). In this paper, we
present optical broad-band and H$\alpha$ imaging of the whole extent of
the disks of the same 12 barred galaxies, and combine parameters derived
from the NIR and optical parts of our imaging survey with parameters
taken from the literature to identify several correlations which
illuminate the important connections that exist between the central
regions and the host galaxy structures at large.

We summarise the observations and data reduction procedures in $\S$~2,
and present our observational results in $\S$~3. Relative \ha\ fluxes
from different parts of the sample galaxies are explored in $\S$~4. In
subsequent sections, we explore correlations between the strength of
the bar and the geometric parameters of the nuclear rings ($\S$~5),
and the shapes of the dust lanes within the bar ($\S$~6). We summarise
our main findings in $\S$~7. Notes on individual objects can be found
in Appendix~1.

\section{Observations and Data Reduction}

\subsection{Optical observations}

\begin{table}
\hfill
\centering
\begin{tabular}{ccccc}
\hline
Galaxy & Date &Telescope&Filters&Source\\ & & & & \\
\hline 
NGC 1300&11/98& JKT &$B$,$I$,H$\alpha$&This paper \\
NGC 1530&11/98& JKT &$B$,H$\alpha$    &This paper \\
        &10/96&     &$I$              &JKT archive\\
NGC 2903&03/98& JKT &$B$,$I$,H$\alpha$&This paper \\
        &11/98&     &$B$,$I$,H$\alpha$&This paper \\
NGC 3351&11/98& JKT &$B$,H$\alpha$    &This paper \\
        &04/99&     &$I$              &JKT archive\\
NGC 3504&11/98& JKT &$B$,H$\alpha$    &This paper \\
        &04/98&     &$I$              &JKT archive\\
NGC 3516&03/98& JKT &$B$,$I$,H$\alpha$&This paper \\
NGC 3982&11/98& JKT &$B$,H$\alpha$    &This paper \\
        &05/88&     &$I$              &JKT archive\\
NGC 4303&11/98& JKT &$B$,H$\alpha$    &This paper \\
        &04/96&     &$I$              &JKT archive\\
NGC 4314&03/98& JKT &$B$,H$\alpha$    &This paper \\
        &11/98&     &$I$              &JKT archive\\
NGC 4321&05/92& INT &$B$,$I$          &KB96\\
        &04/91& WHT &H$\alpha$        &K98\\
NGC 5248&04/95& INT &$B$,$I$,H$\alpha$&This paper \\
NGC 6951&10/95& INT &$B$,$I$          &INT archive\\
        &02/92& WHT &H$\alpha$        &RBK96\\
\hline
\end{tabular}

\caption{Observing details: month of observation (col. 2); telescope
(col. 3), INT is the 2.5m Isaac Newton Telescope, WHT is the 4.2m
William Herschel Telescope; filters (col. 4); and source from which the
image was obtained (col. 5), KB96 is Knapen \& Beckman (1996), RBK96 is
Rozas, Beckman \& Knapen (1996) and K98 is Knapen (1998)} 
\end{table}

We present new and archival broad ($B$ and $I$) and narrow-band
(H$\alpha$) images of the entire bar and disk region of the 12 barred
galaxies that make up our sample. The selection criteria and basic
properties of our sample galaxies (spiral galaxies known to host
star-forming circumnuclear ring-like structure) are given in
Paper~I. The main goal of the present paper is to identify and quantify
basic structural parameters such as the bar length and ellipticity, the
location and shape of dust lanes, and the distribution of massive SF. It
is relatively easy to measure these parameters from multi-band optical
imaging (see, e.g., Knapen et al. 1995a,b; Knapen \& Beckman
1996). Schematically, we use different bands and colours as follows:

\begin{itemize}

\item $I$: to trace the old stellar population in the absence of
significant contamination by young stars and dust. For most galaxies, we
also used $I$-band images as the continuum that is subtracted from the
raw H$\alpha$ images,

\item $B$: in combination with $I$, using $B-I$ colour
index maps, to show the location of dust lanes and sites of SF,

\item H$\alpha$: hydrogen recombination line emission to trace
directly the massive SF.

\end{itemize}

Most images were obtained during two observing runs on the 1.0-m Jacobus
Kapteyn Telescope (JKT), in 1998 March 2-9 and in 1998 November 9-15
(Table~1). We used the TEK4 CCD Camera with a pixel size of 0.331
arcsec/pixel, and a field of view (FOV) on the sky of $5.6\times5.6$
arcmin. The average FWHM seeing is just below 2~arcsec (best seeing is
1.2~arcsec).  The images in $I$ from the first run presented problems
during the flat-fielding process. We have instead used $I$-band data for
these galaxies from the Isaac Newton Group (ING) archive. In addition,
we include several images from the literature, as detailed in Table~1.

\subsection{Data reduction and photometric calibration} 

The images were reduced using standard IRAF tasks. First, the bias level
of the CCD was subtracted from all exposures as a constant after we
verified that the bias level was constant across the chip. The images
were flat-fielded using sky flats taken in each filter at twilight. We
estimated the sky background by determining the mean level on a few
small areas in regions free of galactic emission, although later in the
reduction we checked, and in some cases corrected, these estimates using
the colour profiles (see $\S$ 2.3).  The images were sky-subtracted and
bias-only overscan regions were removed.

After foreground stars were located and their positions determined, we
shifted all the images of each galaxy to a common position. Where
multiple exposures in one band were available, these were median
combined to make the final image of the galaxy in that band. Cosmic-ray
events were automatically removed in the process. Again using the
foreground stars, we calibrated the images astrometrically. The angle of
rotation employed to achieve this was 181.2 degrees counterclockwise for
the images from our two JKT runs.

The photometric calibration of the broad-band images of the galaxies was
done using published aperture photometry where this was available. For
the rest of the galaxies, we used a mean photometric constant. We could
not use standard stars due to unfavourable weather conditions during
both of our JKT runs. Our \ha\ data have not been calibrated for the
same reason.

\subsection{Ellipse fitting}

After calibrating the images, we used the task {\sc ellipse} in IRAF to
fit ellipses to the images, in order to produce radial runs of surface
brightness in the different bands, as well as of parameters such as the
major axis position angle (PA) and the ellipticity of the fitted
ellipses. Unlike in Paper~I, where foreground stars were not a problem
because of the small FOV of the NIR images, here we had to select and
mask out stars that could affect the fitting process. We first
fitted one of the bands ($I$) leaving PA, ellipticity and centre
position as free parameters in the fit. We then imposed the resulting
radial runs of PA, ellipticity and centre position as fixed input values
for the fits to the other two bands. That way, we ensure that when
combining two bands to evaluate a colour ($B$-$I$ in this case), we
combine emission from the same areas in the individual images. $B-I$
radial profiles have the advantage that they are rather sensitive to
uncorrect background subtraction in one of the individual images, which
leads to easily recognizable sharply rising or dropping colour profiles
in the outer regions. This fact allowed us to finetune the background
subtraction, giving more accurate background determinations. We thus
produced radial $B$, $I$ and H$\alpha$ surface brightness, $B-I$, PA and
ellipticity profiles, all of which are shown in Fig.~1.

There was a problem in the practical implementation of this procedure
for a few galaxies, for which the centre is saturated in our $I$-band
image (NGC~4321, NGC~5248 and NGC~6951). We circumvented this problem by
creating a fictitious centre. We located the real position of the centre
from other bands, and added a number of counts to the pixel at this
position (usually no more than 10$\%$ of the surrounding value) using
the IRAF task {\sc pixedit}.

\subsection{Colour index and H$\alpha$ images}

$B-I$ colour maps, which indicate the location of the major dust lanes
in the disk of a galaxy, were produced by dividing one image by the
other. Sky background levels were subtracted before doing this (see
above). Correcting for different seeing values from one band to the
other did not prove necessary. We also produced H$\alpha$ line images,
which are direct tracers of massive SF, by subtracting the continuum
using the $I$-band images. We did not use narrow band continuum filters,
since the use of $I$ as a continuum filter gives sufficiently reliable
results for our mostly morphological purposes, with a considerable
saving in observing time. To determine the scaling factor to be used for
the continuum subtraction in our new images we followed a procedure
based upon the one described by B\"oker et al. (1999). We first found
the pixel by pixel intensity correlation in the original $I$ and
H$\alpha$ images, using {\sc cormap} (a {\sc gipsy} task). Most pixels
in our large images trace background or galaxy continuum, emission from
which which will, after appropriate scaling, be equal in the two
filters. This factor will mostly depend on the shape of the throughput
curves of the filters used. All these pixels will fall along a straight
line, the $x$ and $y$ intercepts of which give an additional estimate
for the sky background in the $I$ and H$\alpha$ images. The minority of
pixels which trace H$\alpha$ line emission will have enhanced H$\alpha$
emission relative to $I$ (continuum), and will be displaced toward
higher H$\alpha$ values in the diagram. Most pixels thus fall on the
easily identifiable straight line indicating the continuum scaling
factor for this particular set of two images (\ha\ line and continuum),
and this factor is determined simply by measuring the slope of the
correlation line. Once we determined the scaling factor, we subtracted
the scaled galactic continuum.

We considered the final subtraction successful when no bumps or hollows
were observed when checking the central region of the final image. More
extensive testing of this method, including a comparison with a method
where fluxes from foreground stars are used to derive the continuum
scaling factor, and a comparison of different filters ($R$, $I$,
narrow-band continuum filter centred at $\lambda=6470$\AA\ and 115\AA\
wide) used for continuum subtraction, will be presented elsewhere
(Knapen et al., in preparation). From that study, and from tests
performed in this paper, we conclude that scaling factors can be
determined accurately as described above, and that resulting
uncertainties in the \ha\ flux level are of the order of 1\% in the disk
to at most 10\% in the \hii\ regions closest to the nucleus. The bulk of
the latter uncertainty originates in colour differences between the
wavelength of the on- and off-line filters, in turn mostly caused by
excess dust extinction in the CNR at the \ha\ wavelength as compared to
that of the $I$ filter.  The continuum subtraction for the H$\alpha$
images of NGC~4321 and NGC~6951 was done differently, and has been
described by Knapen (1998) and Rozas et al.  (1996), respectively.

\subsection{NIR observations}

We obtained broad-band NIR images of the central few kpc regions of our
sample galaxies, as described in detail in Paper~I. The sub-arcsecond
resolution NIR images were obtained with the Canada-France-Hawaii
Telescope. In Paper I, we present the data in the form of sets of
greyscale and contour plots of the broad-band images and of colour index
images derived from them, and sets of radial profiles of surface
brightness, colour, ellipticity and PA.  In Fig.~2, we show overlays of
the central regions of the H$\alpha$ images, described in $\S$~2.1 and
$\S$~2.4, on the NIR $J-K$ colour index images from Paper~I.

\section{Observational results}

\subsection{Morphology of the disks and CNRs}

In Fig.~1, we show for each galaxy greyscale representations of the $I$
broad-band, $B-I$ colour index, and \ha\ narrow-band images, radial
profiles of PA and ellipticity as determined from the fit to the
$I$-band image, and radial surface brightness ($B, I$ and \ha) and
colour ($B-I$) profiles. In the top left corner of each panel of the
figure we show the \ha\ continuum-subtracted emission from the CNR, on
the same scale as the NIR images in Paper~I. Many circumnuclear
features, often in the shape of more or less complete rings (e.g.,
NGC~1300, NGC~3351, NGC~4314, NGC~6951) or spiral arm-like features
(e.g., NGC~1530, NGC~4321) can easily be distinguished. Except in
NGC~3516 and NGC~4314, \ha\ emission from the spiral arms in the disk is
obvious. Some of the bars (e.g., those in NGC~1530, NGC~3504, NGC~5248)
show clear signs of massive SF, whereas others (e.g., those in NGC~1300,
NGC~3351, NGC~4303, NGC~6951) are completely devoid of
\ha\ emission.  The $B-I$ colour index maps show (in lighter tones,
indicating bluer colours) predominantly where SF occurs at present,
mostly in the CNR and the disk spiral arms, but also (in darker tones,
indicating redder colours) where the main dust lanes are
located. While organised dust lanes are often seen outlining spiral
arms in the disk, we will in the present paper concentrate on the dust
lanes in the bars, which are thought to outline the locations of
shocked gas, and are thus prime tracers of the bar dynamics (see
$\S$~6).

In Fig.~2, we present overlays of the \ha\ emission (in contours) from
the CNR on the $J-K$ colour index images from Paper~I for all sample
galaxies.  In general, the location of the nuclear ring in H$\alpha$
corresponds to that in the NIR, but in several cases the \ha\ is more
extended (e.g., NGC~1300, NGC~3516, NGC~3982, NGC~4303). More specific
details on individual galaxies are summarised in Appendix~1.

\subsection{Bar and CNR parameters}

\begin{table*}
\centering
\begin{tabular}{ccccccccc}
\hline
Galaxy & Scale&Morphological & Nuclear & CNR & Disk scale & Disk PA &
Inclination\\
       & pc/arcsec&type  & activity & feature & length (kpc) & (\deg) &
angle (\deg)\\
\hline 
NGC 1300 & 101 &(R')SB(s)bc & --&Ring               & 21.5 & 87 & 35\\
NGC 1530 & 159 &SB(rs)b     & --&Ring+spiral        &  9.6 & 8 & 45\\
NGC 2903 &  36 &SAB(rs)bc   & Starburst&Ring        &  3.4 & 22 & 65\\
NGC 3351 &  49 &SB(r)b      & Starburst&Ring        &  1.9 & 13 & 40\\
NGC 3504 &  99 &(R)SAB(s)ab & Starburst&Ring+spiral &  2.0 & 147 & 22\\
NGC 3516 & 171 & (R)SB(s)    & Sy 1.5&Ring          &  4.2 & 46 & 34\\
NGC 3982 &  72 &SAB(r)b     & Sy 2&Spiral           &  0.6 & 6 & 14\\
NGC 4303 & 101 &SAB(rs)a    & Sy 2&Ring             &  5.7 & 116 & 22\\
NGC 4314 &  62 &SB(rs)a     & LINER&Ring+spiral     &  3.2 & 59 & 27\\
NGC 4321 &  70 &SAB(s)bc    & --&Ring+spiral        &  3.1 & 153 & 30\\
NGC 5248 &  74 &SAB(rs)bc   & --&Ring+spiral        &  3.3 & 105 & 40\\
NGC 6951 &  92 &SAB(rs)bc   & LINER/Sy &Ring+spiral &  3.7 & 157 & 44\\
\hline
\end{tabular}
\caption{Properties of the sample galaxies. NGC number (col. 1), scale
(col. 2, Paper~I); morphological type (col. 3, from de Vaucouleurs et
al. 1991, hereafter RC3), presence and type of nuclear activity (col. 4,
from NED; Sy is Seyfert), features in the CNR (col.~5, Paper~I), scale
length of the exponential disk (col.~6), and deprojection parameters: PA
of the major axis of the outer disk (col.~7) and galaxy inclination
(col.~8).}

\end{table*}

\begin{table*}
\centering
\begin{tabular}{cccccccccc}
\hline
Galaxy & Dimensions of & Ring & Ring $\epsilon$ & PA offset (\deg) & Bar Length& \multicolumn{2}{c}{Bar ellipticity} & Bar
strength & $\Delta \alpha$ \\
       & \ha\ ring (kpc) & size & Deproj. & Deproj. & (kpc) & $\epsilon$ & $\epsilon_{\rm d}$ & $Q_{\rm b}$ & (\deg/kpc)\\
\hline 
NGC 1300  & 0.5$\times$0.5 & 0.026 & 0.18 & 28 & 8.6 &0.72 & 0.58   & 0.44 & 5\\
NGC 1530  & 0.7$\times$0.6 & 0.032 & 0.19 & 85 & 11.0 & 0.61 & 0.65   & 0.71 & 3\\
NGC 2903  & 0.4$\times$0.2 & 0.030 & 0.17 & 18 & 2.7 & 0.60 & 0.15 & 0.26 & - \\
NGC 3351  & 0.7$\times$0.6 & 0.064 & 0.11 & 88 & 3.1 & 0.39 & 0.58   & 0.20 & 12 \\
NGC 3504  & 0.4$\times$0.3 & 0.050 & 0.10 & 43 & 2.9 & 0.62 & 0.36 & 0.26 & 11 \\
NGC 3516  &  ---  & -- & -- & -- & 2.4 & 0.30 & 0.39    & - & - \\
NGC 3982  &  ---  & -- & -- & -- & 0.8 & 0.41 & 0.37  & - & - \\
NGC 4303  & 0.6$\times$0.6 & 0.030 & 0.07 & 53 & 4.3 & 0.54 & 0.58  & 0.27 & 9 \\
NGC 4314  & 0.5$\times$0.4 & 0.064 & 0.27 & 12 & 4.8 & 0.63 & 0.67   & 0.33 & 7 \\
NGC 4321  & 0.9$\times$0.6 & 0.058 & 0.26 & 19 & 3.7 & 0.50 & 0.50 & 0.20 & 18 \\
NGC 5248  & 0.9$\times$0.9 & 0.066 & 0.23 & 10 & 7.8 & 0.48 & 0.51 & 0.03 & 20 \\
NGC 6951  & 0.5$\times$0.3 & 0.046 & 0.29 & 46 & 5.5 & 0.50 & 0.52 & 0.39 & 0 \\
\hline
\end{tabular}
\caption{Properties of the nuclear rings and the bars in our sample
galaxies: measured dimensions of the nuclear ring (major and minor
axis radii, in kpc, col.~2), ring size expressed as $D_{\rm
ring}/D_{25}$, where $D_{25}$ is from the RC3 (Col.~3), deprojected
ring ellipticity (col.~4), deprojected (col.~5) PA offset (in degrees)
between major axes of the nuclear ring and of the bar; deprojected bar
length in kpc (col.~6); bar ellipticity as determined from our data,
as measured (col.~7) and deprojected (col.~8); bar strength $Q_{\rm
b}$ (col.~9; from Block et al. 2001, E.~ Laurikainen \& H.~Salo, in
preparation, and R. Buta, private communication; the average is given
where a galaxy is included in more than one study); and dust lane
curvature measure $\Delta \alpha$ (see text, col.~10).}

\end{table*}

We summarise selected properties of the sample galaxies in Tables~2
and~3, as taken from the literature and from Paper I, and as derived in
this paper.  Table~2 lists the scale in parsec/arcsec, the morphological
type, type of nuclear activity, and a brief description of the observed
CNR features for all sample galaxies. We determined the scale length of
the exponential disk of each galaxy by least-squares fitting of the
relevant part of the $I$-band surface brightness profile (Fig.~1).  The
dimensions of the nuclear ring were measured directly from the images,
and are given in Table~3 as major and minor axis sizes, in kpc. The size
of the ring was measured by tracing the ridge line of the ring from the
colour index maps (for the NIR) or the \ha\ images (for the SF), and
fitting these coordinates to an ellipse to find the size and the
ellipticity. Allowing for the different kinds of imaging used, our ring
dimension measurements generally agree rather well with those published
by Buta \& Crocker (1993) for the ten galaxies common to their and our
samples. We also derived the relative size and deprojected (or
intrinsic) ellipticity of the nuclear rings, which are listed in
Table~3. Typical uncertainties in these quantities are 10\% and 0.05,
respectively. The relative size of the ring was defined as the diameter
of the ring divided by the diameter of the disk, where we use $D_{25}$
from the RC3 to find the latter value. Laine et al. (2002) showed that
the distribution of these relative nuclear ring sizes peaks at $r_{\rm
ring}/D_{25}=0.06$.  The values found here are in agreement with their
result.

We used the NIR and optical radial profiles in combination with the
images to measure the difference between the PAs of the nuclear ring and
of the bar. Using the disk parameters given in Table~2 we deprojected
this angle. The final PA offset between the bar and the ring is given in
Table~3. Typical uncertainties, resulting from small uncertainties in
the PAs as measured from the profiles and images and in the deprojection
parameters, are $\pm7\deg$.

The basic bar parameters length and ellipticity are also tabulated in
Table~3. We define the end of the bar as the radius where the
ellipticity reaches its local maximum value.  The uncertainty in the bar
length is on the order of 5\% -- 10\%. The bulge component usually has a
radius of only 1 kpc or so, and is not expected to affect the measured
bar lengths. We used the scales as given in Table~2 to calculate the
values in kpc, and the disk parameters (Table~2) to deproject the bar
ellipticities and lengths.

To evaluate the ellipticities of the bars, often, but strictly speaking
incorrectly, referred to as bar strengths, we followed the simple
procedure outlined by Martin (1995), and implemented by many other
authors (e.g., Martinet \& Friedli 1997; Chapelon et al. 1999; Knapen et
al. 2000a).  Considering bars to be ellipsoidal features, their shape
can be described by the ellipticity of the bar $\epsilon = 1 -
b/a$(where $b/a$ is the bar axis ratio). A value of $\epsilon = 0$
corresponds to the absence of a bar, whereas $\epsilon = 0.8$ denotes
the thinnest class of bars. Following the approach of, among others,
Knapen et al. (2000a), we have determined the bar ellipticity from the
maximum in the radial ellipticity profile in a region of roughly
constant PA.  This value is then deprojected using the (disk)
ellipticity at large radii in the disk of the galaxy, and the difference
between the bar PA and the PA of the major axis of the outer disk. The
results are listed in Table~3, where typical uncertainties in the
ellipticity measures are 0.05.

Recently, bar strength has been addressed in a more accurate sense on
the basis of NIR imaging, basically by measuring the ratio of
non-axisymmetric to axisymmetric forces (e.g., Buta \& Block 2001;
Block et al. 2001; Laurikainen, Salo \& Rautiainen
2002). Determinations of the bar strength parameter $Q_{\rm b}$ by
Block et al. and Laurikainen et al. in general agree well
(E. Laurikainen, private communication), but only broad agreement
between this bar strength and the bar ellipticity has been found
(Block et al. 2001). In Table~3, we also list values for $Q_{\rm b}$
for those galaxies where this has been determined. Typical errors in
$Q_{\rm b}$ are $\pm0.1$.

\setcounter{figure}{2}

\begin{figure}
\psfig{{figure=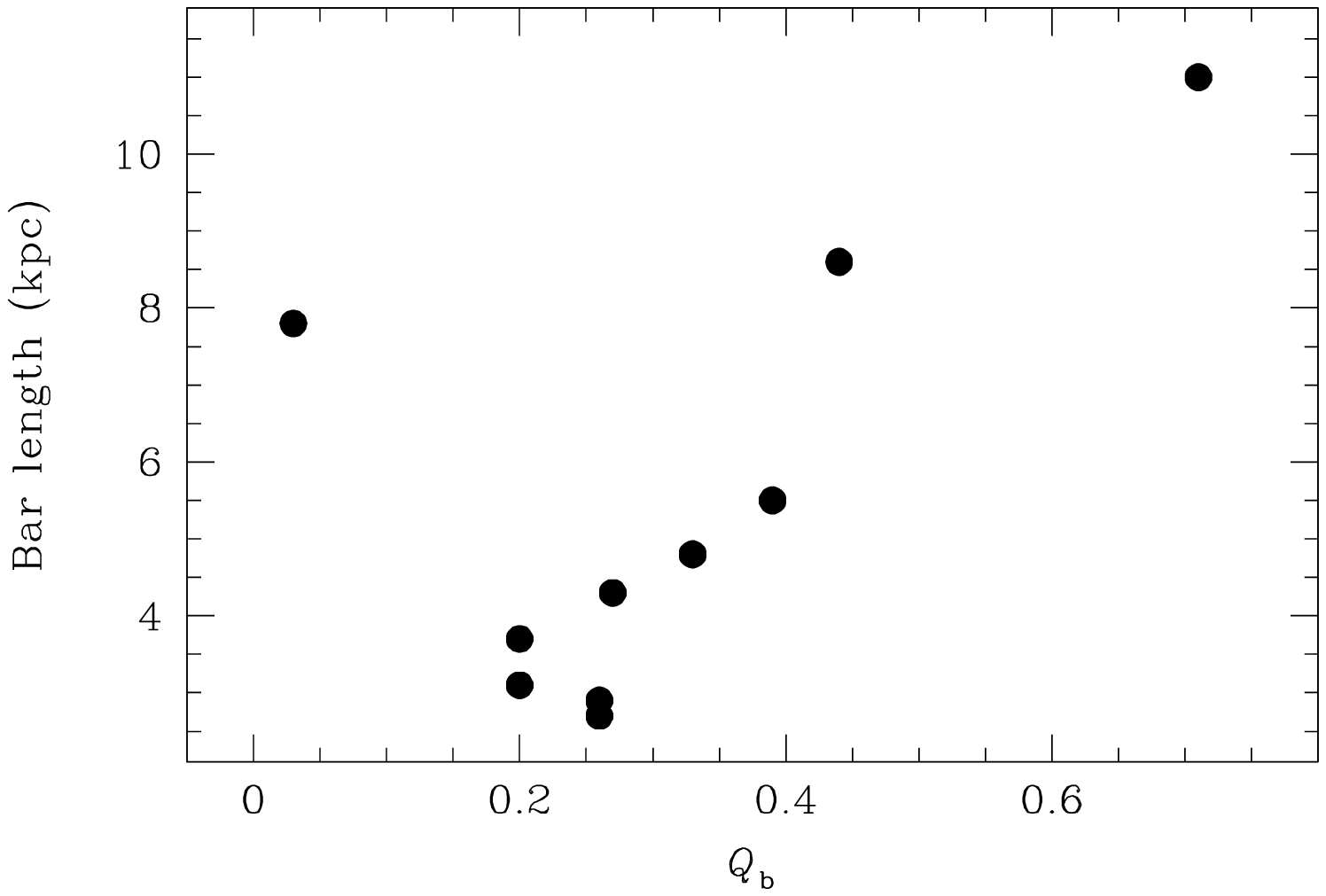},height=6cm}
\caption{{\bf a.}Bar length in kpc as function of bar strength. Typical
uncertainties are 5-10\% in bar length and 0.1 in $Q_{\rm b}$.}
\end{figure}

\setcounter{figure}{2}

\begin{figure}
\psfig{{figure=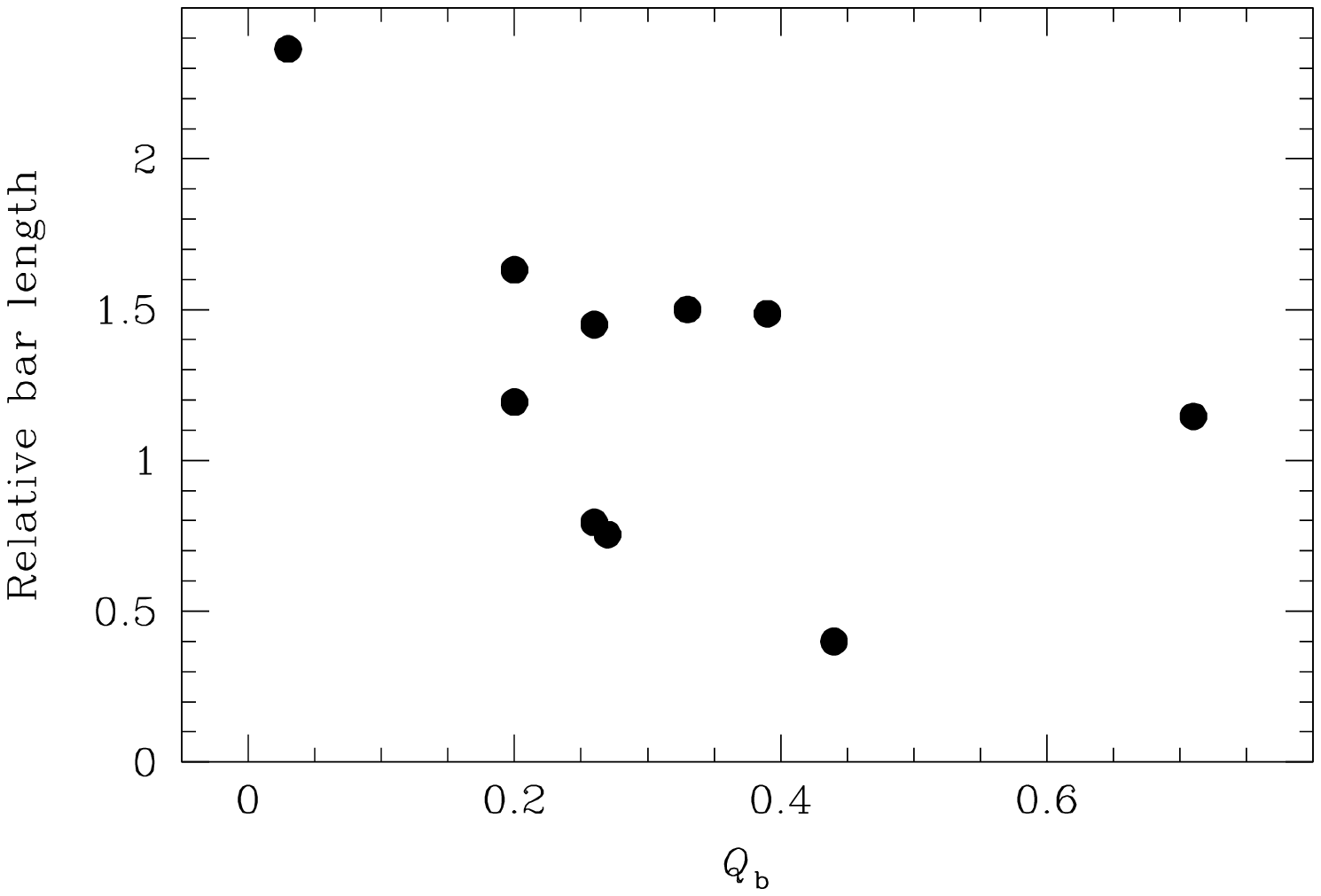},height=6cm}
\caption{{\bf b.}Relative bar length, or the ratio of bar length and disk
exponential scale length, as function of bar strength. Typical
uncertainties are 10-15\% in relative bar length and 0.1 in $Q_{\rm b}$.}
\end{figure}
\setcounter{figure}{2}

\begin{figure}
\psfig{{figure=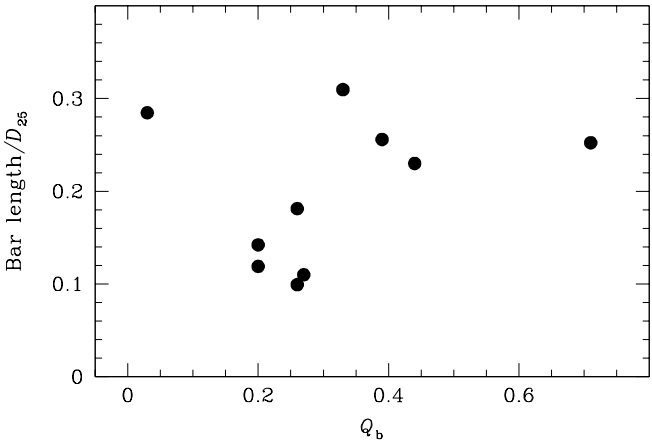},height=6cm}
\caption{{\bf c.}As Fig.~2b, but now for the ratio of bar length and
galaxy size. The RC3 parameter $D_{25}$ was used for the latter value.}
\end{figure}

With the information in hand, we can easily check the correlation
between the bar ``strength'' and length, as reported by Martinet \&
Friedli (1997). In Fig.~3a, we show how the length of the bar is
correlated with the strength of the bar, as measured through $Q_{\rm
b}$. This figure at first sight indeed confirms the claim by Martinet \&
Friedli that stronger bars are longer. But if instead of the bar length
in kpc the length of the bar relative to the scale length of the
exponential disk of its host galaxy is plotted (Fig.~3b), the
correlation disappears, and there is even a hint that shorter bars are
stronger, which is opposite to the effect reported by Martinet \&
Friedli (1997). Reproducing the measure used by Martinet \& Friedli, bar
length relative to the size of the host galaxy ($D_{25}$ from the RC3),
we find no correlation at all (Fig.~3c). Reasons for the confusing set
of results may include the small sample size in our study, but also the
use by Martinet \& Friedli of parameters measured by Martin (1995), who
derived bar lengths and ellipticities from optical (as opposed to NIR)
images, and who used bar ellipticity as a measure of bar
strength. Proper analysis of a larger galaxy sample is needed to confirm
any relation between lengths and strengths of galactic bars. We draw
attention to a deviant point in Fig.~3a,b, which denotes the very weak
bar in NGC~5248 ($Q_{\rm b}=0.03$). This may well be a special case, as
described in detail by Jogee et al. (2002a,b).

\subsection{Radial profiles}

Some of the most important morphological characteristics of each galaxy
can be recognised in radial profiles such as the ones shown in Fig.~1.
In NGC~4321, Knapen \& Beckman (1996) identified four main regions in
the H$\alpha$ profile, reflected also in, e.g., the $B-I$ profile: the
nuclear area corresponding to a central peak and explained by the
existence of enhanced SF in and around the nucleus; the bar region with
a relatively smaller SF rate, seen as a dip in the profile; a region
where the H$\alpha$ luminosity falls off exponentially, which
corresponds to the star forming spiral arms in the disk; and finally an
outer region of steeper decline. From their study of \hii\ region
distributions in spiral galaxies, Gonz\'alez Delgado \& P\'erez (1997)
note that the barred galaxies in their sample tend to have either rising
or flat central parts in the radial distribution of the number of \hii\
regions per unit surface area, which is probably connected to the lack
of SF in the bar. It is interesting in this respect that the one
exception mentioned by Gonz\'alez Delgado \& P\'erez (1997) is NGC~7479,
well-known for being a strongly barred galaxy with exceptionally
prominent SF within the bar, a bar which may have been created in a
recent minor merger (Laine \& Heller 1999).

We can now try to generalise findings such as those by Knapen \& Beckman
(1996) using the 12 H$\alpha$ radial profiles for the barred galaxies in
our sample. Considering primarily the \ha\ profiles, we find that most
galaxies have enhanced emission in the CNR, although in some cases
non-stellar emission from the nucleus may contribute to a central rise
in the profile (e.g., in NGC~3516). Very often the profile shows a dip
in the region of the bar. Particularly good examples are NGC~1300,
NGC~3351 and NGC~6951.  In a number of galaxies, such as in NGC~1530,
the profile shows a small rise corresponding to zones of enhanced SF
near the ends of the bar, clearly visible in the images.  The inner
rings near the ends of the bar in NGC~3351 and NGC~3504 can be
recognised as bumps in the H$\alpha$ profile.  A few galaxies (e.g.,
NGC~4303 and NGC~6951) clearly show the SF in the disk outside the bar
region, just as in NGC~4321 (Knapen \& Beckman 1996). In the case of
NGC~2903 there is a dominant continuous component in the profile, and
individual star forming regions are hard to identify. We can conclude
that the radial zones identified by Knapen \& Beckman (1996) in their
study of NGC~4321 are not all present in each of the galaxies studied
here. However, depending on the distribution of SF in the galaxies,
morphologically important regions can be recognised in the profiles,
such as the CNR with enhanced, or the bar with depressed SF. In general,
and as expected, the results obtained from the radial profiles are
consistent with the morphology as summarised in Appendix~1.

Where specific zones are seen in the H$\alpha$ profile, these can
usually be recognised in the $B$ and $I$ profiles as well. These
profiles behave in a very similar way radially: central peak;
depression in the bar region, less pronounced in $I$, as expected for
an older, redder, bar population; inner ring where present; and
exponential disk.  The $B-I$ colour index profile also follows the
large-scale distribution of SF in most galaxies, showing a blue peak
in the CNR and a red depression in the bar zone.

\section{Relative H$\alpha$ flux from bars and circumnuclear regions}

\begin{table*}
\centering
\begin{tabular}{cccc}
\hline
Galaxy & \% Flux from the disk&\% Flux from the bar&\% Flux from the CNR \\
& (excluding bar and CNR)&(excluding CNR)\% &(excluding nucleus)\\
\hline 
NGC 1300 &67&24  & 9   \\
NGC 1530 &31&37  &29   \\
NGC 2903 &68&19  &13   \\
NGC 3351 &51&10  &38   \\
NGC 3504 & 6&20  &66   \\
NGC 3516 &11& 0  &45   \\
NGC 3982 &91& 0  & 0   \\
NGC 4303 &69&19  &12   \\
NGC 4314 & 0&17  &74   \\
NGC 4321 &73&11  &16   \\
NGC 5248 &63&14  &20   \\
NGC 6951 &77& 7  &15   \\
\hline
\end{tabular}
\caption{Relative flux from the disk, bar and circumnuclear
regions. The latter does not contain the flux from the nuclear point
source, if present, which is also why the percentages given here do
not add up to 100\% in all galaxies. Typical uncertainties are 25\%,
10\% and a few to 10\% for disk, bar and CNR values, respectively (see
text).}
\end{table*}

In order to study whether and how the large-scale parameters of disks of
galaxies influence the relative distribution of the massive SF within
those disks, we have measured the relative H$\alpha$ flux from the bars
and the CNRs. We have done this by separating the images into different
regions: the CNR, the bar and the disk, and determining the flux from
these areas with the aperture photometry package within {\sc gaia}. The
separation of the \ha\ images into individual areas was done by
carefully judging where the CNR and bar regions end, which is in general
very easy to recognise in \ha\ (see Fig.~1), but also referring to the
$I$ and $B-I$ images in doing so. While this may well introduce a small
uncertainty in our measurements, the main source of uncertainty,
especially for the entire disk, is the determination of the background
sky value. The resulting uncertainties are of the order of 25$\%$ in the
flux from the total disk, 10$\%$ in the case of the bar, and one to a
few \% in the CNRs. Given the uncertainties in the continuum subtraction
in the nuclear regions (see \S2.4), we conservatively increase the error
margin for the CNR flux ratio to 10\%.

For NGC~3516 and NGC~3982 no bar could be defined, and no bar flux
values for these galaxies are included in the table. In order to take
out possible disturbing effects due to \ha\ emission from active nuclei,
we subtracted the flux from a nuclear point source, where present, from
the CNR flux. Only in one galaxy (NGC~3516) is the nuclear
\ha\ flux fraction high (44\%) and hard to determine, and we do not
consider this galaxy in the analysis below. Of the three galaxies with
nuclear flux fractions of 8\%-9\%, in two the nuclear point source is easy
to identify (NGC~3982 and NGC~4314, see Fig.~1g,i), whereas in NGC~3504
the uncertainty is slightly higher. In all other galaxies, the nuclear
\ha\ flux fraction is zero or low (0-3\%) and relatively easy to determine.
Table~4 lists the relative fluxes thus obtained for the CNR, bar, and
disk, for our sample galaxies.

In certain sample galaxies (NGC~3504 and NGC~4314) the \ha\ flux from
the CNR dominates the total \ha\ emission from the host galaxy. The
relative contribution of the bar itself is substantial in galaxies such
as NGC~1300 and NGC~1530, but is generally at the level of around 10$\%$
of the total flux. The contribution from the disk is very small or even
negligible in NGC~3504, NGC~3516 and NGC~4314, substantial or dominant
in all others.

\begin{figure}
\centerline{
\psfig{{figure=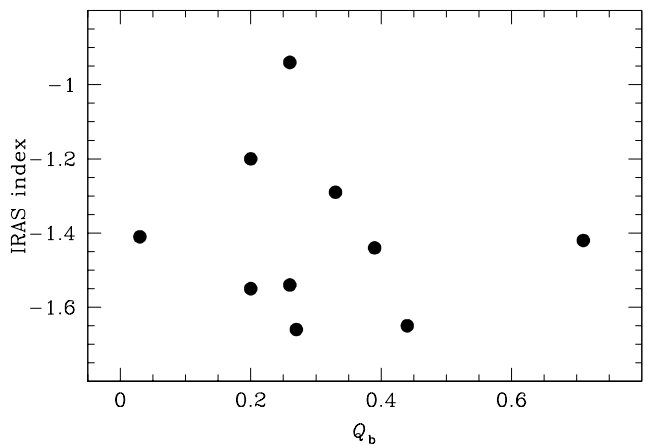},height=6cm}
}
\caption{IRAS index log($S_{25}$/$S_{100}$) as a function of the
strength of the bar, $Q_{\rm b}$.}
\end{figure}

\begin{figure}
\centerline{
\psfig{{figure=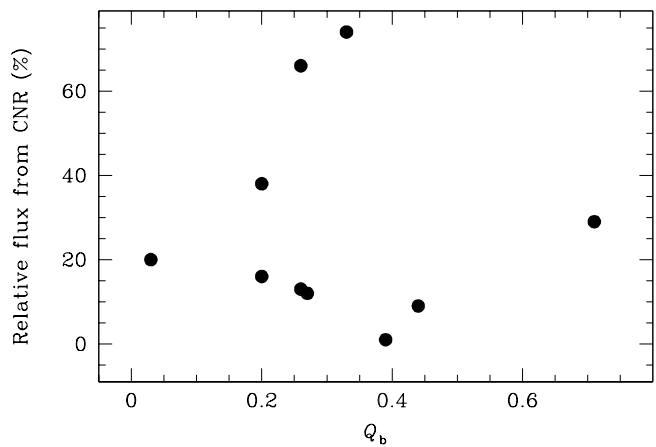},height=6cm}
}
\caption{Relative flux from the CNR (excluding the nucleus), plotted as
a function of the strength of the bar, $Q_{\rm b}$.}
\end{figure}

\begin{figure}
\centerline{
\psfig{{figure=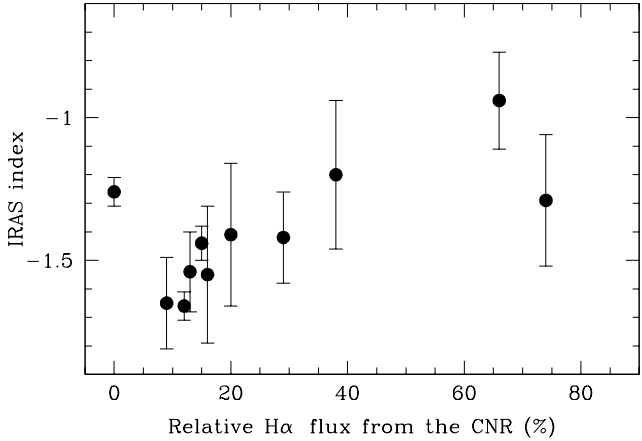},height=6cm}
}
\caption{Relative flux from the CNR (excluding the nucleus) vs. the IRAS
index log($S_{25}$/$S_{100}$). Typical errors along the $x$-axis are a
few to ten percent (see text).}
\end{figure}

Various authors have used far-infrared (FIR) colour indices as direct
estimators of current SF activity. Puxley, Hawarden \& Mountain (1988)
established a value of log($S_{12}$/$S_{25}) < - 0.35$ as a signature
of the existence of regions of SF.  Dultzin-Hacyan, Moles \& Masegosa
(1988) studied a sample of \hii\ galaxies and discussed the use of
log($S_{25}$/$S_{100}$) to evaluate the SF activity, giving median
values of log($S_{25}$/$S_{100}$) $=-$1.4 for normal galaxies, $-$1.15
for LINERs, $-$0.8 for starburst and Seyfert~2 galaxies, and $-0.5$
for Seyfert~1 galaxies. Martinet \& Friedli (1997), using the same
parameter, considered two classes of FIR colours
(log($S_{25}$/$S_{100}$)$\geq$$-$1.2 and
log($S_{25}$/$S_{100}$)$<-$1.2) corresponding to more or less
pronounced SF activity, respectively.

\begin{table*}
\centering
\begin{tabular}{cccccc}
\hline
Galaxy & $S_{25}$ & $\%$ & $S_{100}$ & $\%$ & log($S_{25}$/$S_{100}$)\\ 
       & (W m$^{-2}$ Hz$^{-1})$& error& (W m$^{-2}$ Hz$^{-1}$) & error& \\
\hline 
NGC 1300 & 2.30E$-$27  &  0 & 1.03E$-$25 & 10 & $-$1.65$\pm$0.16\\
NGC 1530 & 8.40E$-$27  &  7 & 2.19E$-$25 &  9 & $-$1.42$\pm$0.16\\
NGC 2903 & 2.90E$-$26  &  6 & 1.01E$-$24 &  7 & $-$1.54$\pm$0.14\\
NGC 3351 & 2.09E$-$26  & 12 & 3.35E$-$25 & 18 & $-$1.20$\pm$0.26\\
NGC 3504 & 3.73E$-$26  &  9 & 3.27E$-$25 & 16 & $-$0.94$\pm$0.17\\
NGC 3516 & 8.94E$-$27  &  7 & 2.26E$-$26 &  6 & $-$0.40$\pm$0.04\\
NGC 3982 & 8.33E$-$27  &  3 & 1.52E$-$25 &  3 & $-$1.26$\pm$0.05\\
NGC 4303 & 1.40E$-$26  &  1 & 6.47E$-$25 &  3 & $-$1.66$\pm$0.05\\
NGC 4314 & 3.62E$-$27  & 12 & 7.14E$-$26 & 13 & $-$1.29$\pm$0.23\\
NGC 4321 & 1.57E$-$26  & 11 & 5.62E$-$25 & 11 & $-$1.55$\pm$0.24\\
NGC 5248 & 1.74E$-$26  & 12 & 4.45E$-$25 & 13 & $-$1.41$\pm$0.25\\
NGC 6951 & 1.37E$-$26  &  3 & 3.75E$-$25 &  3 & $-$1.44$\pm$0.06\\
\hline
\end{tabular}
\caption{IRAS flux values for our sample galaxies, as obtained from
Moshir et al. (1990)}
\end{table*}

We have compiled IRAS fluxes for our sample galaxies from the literature
(Table~5). We explored possible relations between the IRAS index
log($S_{25}$/$S_{100}$), used as an indicative value to quantify the
global SF, and the length and strength of the bars, but found no trends
from our small sample of galaxies (e.g., Fig.~4 shows the IRAS index
vs. bar strength $Q_{\rm b}$). From larger samples of galaxies, but
using less reliable methodology, Martinet \& Friedli (1997) and Aguerri
(1999) found that galaxies with enhanced global SF as measured from
their IRAS fluxes tend to have longer and more elliptical
bars\footnote{Note that both these studies used bar parameters
determined from optical imaging, and that Aguerri did not de-project
these parameters.}.  Equally, we cannot distinguish any trends in the
distribution of SF as a function of bar strength $Q_{\rm b}$ (as an
example, we show the run of the relative flux from the CNR against bar
strength in Fig.~5) or length (not shown) of the bar. A related
conclusion is drawn by Sheth (2001) who finds that the ratio of nuclear
over disk molecular gas surface density does {\it not} correlate with
the length of the bar. This ratio itself is much enhanced in barred with
respect to non-barred galaxies (see also Sakamoto et al. 1999) .

Figure~6 presents the run of the relative flux from the CNR against the
IRAS log($S_{25}$/$S_{100}$) index, quantifying the global SF. It shows
that a higher \ha\ flux fraction from the CNR leads to higher IRAS
indices. This may be due to the IRAS index being more sensitive to CNR
SF, which is concentrated in a more dusty environment than disk SF. It
may also imply, though, that the more centrally concentrated the star
formation is, the more SF takes place in absolute terms in the whole of
the host galaxy.  Although we subtracted the \ha\ flux which might be a
result of non-stellar nuclear activity rather than circumnuclear SF, a
caveat here is the role played by nuclear activity in the FIR emission,
which we cannot quantify for our sample galaxies in the absence of FIR
spectroscopy. On the basis of studies of other galaxies (see e.g. Genzel
\& Cesarsky 2000), though, we estimate that the Seyfert nucleus
contributes at most 20\%, and possibly as little as 5\%, to the total
flux, with the exception of NGC~3516. For this galaxy, already excluded
from our analysis on other grounds, Mouri \& Tanaguchi (2002) claim that
the FIR luminosity is dominated by the AGN contribution. Only the study
of a larger sample of galaxies can confirm the tentative relation we
report here. We have started to undertake such a larger study.

\section{Orientation of bars and nuclear rings}

\begin{figure}
\centerline{\psfig{{figure=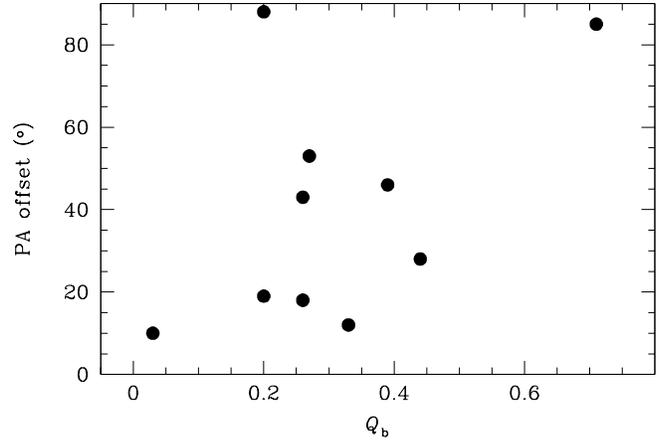},height=6cm}}
\caption{Difference between the major axis PAs of the
nuclear ring and the bar which hosts it, as a function of bar strength
$Q_{\rm b}$. Typical uncertainties are 7\% in PA offset
and 0.1 in $Q_{\rm b}$. }
\end{figure}

\begin{figure}
\centerline{\psfig{{figure=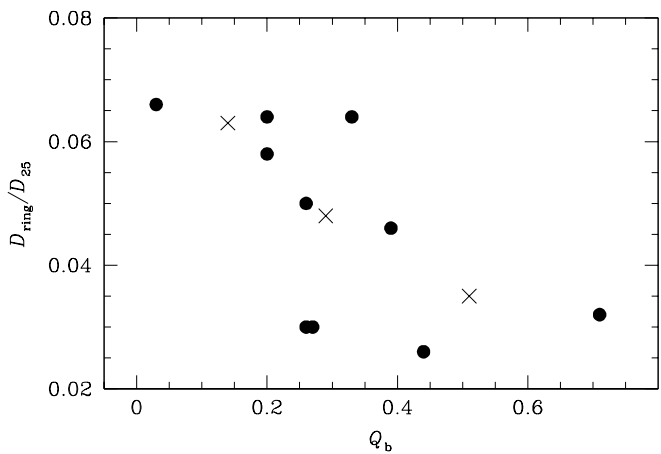},height=6cm}}
\caption{Relative size of the nuclear ring (ring diameter divided by 
$D_{25}$ of the galaxy) as a function of bar strength $Q_{\rm b}$
(filled dots). Typical uncertainties are 10\% in relative ring size and
0.1 in $Q_{\rm b}$. Crosses indicate averages of the data points with
$Q_{\rm b}$ values between 0 and 0.25 (3 galaxies), 0.26 and 0.35 (4),
and 0.39 and 1 (3), respectively.}
\end{figure}

\begin{figure}
\centerline{\psfig{{figure=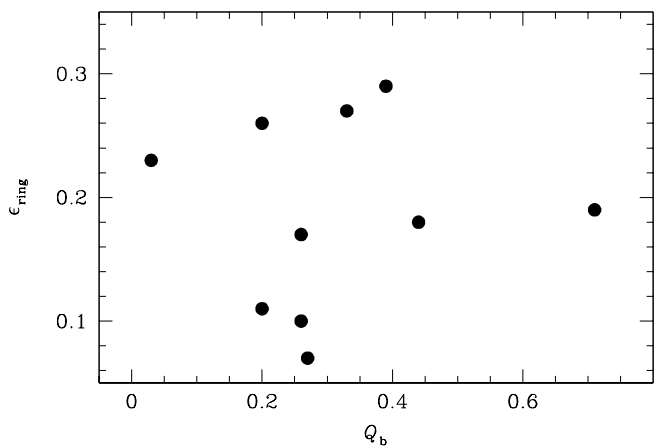},height=6cm}}
\caption{Deprojected, or intrinsic, ellipticity of the nuclear ring
as a function of bar strength $Q_{\rm b}$. Typical
uncertainties are 0.05 in ring $\epsilon$ and 0.1 in $Q_{\rm b}$.}
\end{figure}

From our H$\alpha$ images and radial profiles we have measured the
sizes of the nuclear rings, the deprojected differences between the
PAs of the major axes of the rings and of the large-scale bar of their
host galaxies, as well as the deprojected, or intrinsic, ellipticities
of the rings (Table~3). Since for most of our galaxies we also know
the bar strength, we can now, for the first time, explore
observationally how these geometric ring parameters depend on the
strength of the bar potential that hosts them. 

The relative orientations of the nuclear rings span the complete range
from aligned to perpendicular to the bar, with no preferred PA
offset. This was also found by Buta \& Crocker (1993). These authors
describe a number of caveats in the derivation of PA offsets which also
apply to our work, namely selection effects, sample size, and difficulty
of properly defining ring shapes and orientation. In Fig.~7, we plot the
observed PA offset between the nuclear ring and the bar as a function of
the bar strength $Q_{\rm b}$. Even though the galaxy with the weakest
bar (NGC~5248) has the smallest PA offset and the one with the strongest
bar (NGC~1530) one of the largest, there is no obvious correlation.

Fig.~8 shows the run of the relative size of the nuclear ring, defined
as the ring diameter divided by the host galaxy diameter, as a
function of bar strength $Q_{\rm b}$. It is seen that stronger bars
host smaller rings. Finally, Fig.~9 shows that the intrinsic
ellipticity of the rings does not vary with different bar strength.

Numerical modelling (e.g., Knapen et al. 1995b; Heller \& Shlosman 1996)
gives important clues to the evolution of the geometric properties of a
nuclear ring which forms near the ILRs in a barred galaxy. The first
point to note is that stronger, thinner, bars lead to smaller nuclear
rings, exactly as we see here in Fig.~8. This can be seen, e.g., in
fig.~11 of Knapen et al. (1995b), from the form and extent of the
representative stellar periodic orbits in a bar with ILRs. The main
orbit family, that of the so-called $x1$ orbits (see Contopoulos \&
Papayannopoulos 1980), forms the backbone of the stellar bar. A second
orbit family, that of the $x2$ orbits, is oriented perpendicularly to
the $x1$ orbits in the region of the ILRs. Since nuclear rings are
gaseous features, it is important to note that gas can only populate
those regions where the $x1$ and $x2$ orbits are not intersecting, which
is the inner part of the galaxy. Fig.~11 of Knapen et al. (1995b) makes
clear that, in order to avoid the intersecting orbits, gas will have to
settle further in in the case of stronger, or thinner, bars.  This then
leads to smaller areas where the nuclear ring can form, and thus to
smaller nuclear rings. Of course smaller nuclear rings in this context
is a relative term, and we have chosen here to use the size of the ring
relative to the total size of the galaxy (following Laine et
al. 2002). However, if we use the size of the ring relative to the
length of the bar as a measure, we find a similar correlation (not
shown) as depicted in Fig.~8.

The situation is different for both the PA offsets from the bar, and the
intrinsic ellipticities, of the nuclear rings. As described by, e.g.,
Knapen et al. (1995b) and Shlosman (1999), these parameters are
regulated primarily by the gas inflow into the CNR. As can be seen in,
e.g., fig.~13 of Knapen et al. (1995b), less inflow will lead to a
smaller leading angle of the ring with respect to the bar, and to a
smaller ellipticity of the ring: it becomes rounder. Since gas inflow
does not depend exclusively on the parameters of the bar, such as its
strength, but more generally on the availability of fuel, correlations
of ring ellipticity and PA offset are not expected, and in fact not
observed (Figs.~7, 9).  The dynamical timescale in the CNR is of the
order of $10^7$ yrs, and thus the availability of the fuel can be
expected to be variable, further eliminating any correlation with, e.g.,
the bar strength.  Because the dynamical timescale of a large stellar
bar is much longer ($\sim10^8$ yrs) the size of a nuclear ring will be
much more stable in time, given the arguments outlined in the preceding
paragraph. This picture is nicely illustrated by the recent results of
Jogee et al. (2002b) for the nuclear ring of NGC~5248. They found that
the younger population in the ring, as traced by FUV emission, outlines
spiral arms and is more elliptical in shape, whereas the relaxed ring,
as seen in older stellar population, is much rounder.

\section{Bars and dust lanes within them}

Colour index maps reveal interesting symmetric dust lane patterns in the
bars of strongly and weakly barred galaxies. The curvature of the dust
lanes has theoretically been related to the strength of the bar, in the
sense that weak bars are associated with curved dust lanes, whereas
strong bars have straight dust lanes (Athanassoula 1992b).

We have used a simple method to measure and quantify the curvature of
the dust lanes, in an attempt to probe the anti-correlation
theoretically predicted by Athanassoula (1992b) between the degree of
curvature of the bar dust lanes and the strength of the bar. From
deprojected $B-I$ and $B$-band images, where the dust lanes are well
visible, we have measured the change in the angle of the tangent to
the curved dust lane, and express the result $\Delta \alpha$ in units
of degrees per kiloparsec distance along the dust lane. We ignore those
parts of the dust lanes (e.g., where the lanes connect to the CNR)
where the curvature increases or decreases sharply. Well-defined,
twofold symmetric dust lanes could not be identified clearly enough in
NGC~2903, NGC~3516, and NGC~3982, and these galaxies are not
considered in the analysis below.

\begin{figure}
\psfig{{figure=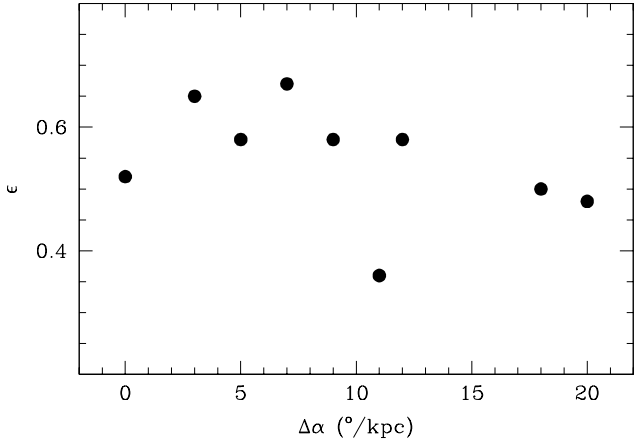},height=6cm}
\caption{Deprojected ellipticity of the bar as a function of dust lane
curvature index $\Delta \alpha$. Straight dust lanes in the bar are on
the left side of the diagram, and thin bars are towards the top. Typical
uncertainties are 0.05 in bar $\epsilon$ and 3\deg/kpc in $\Delta\alpha$.}
\end{figure}

\begin{figure}
\psfig{{figure=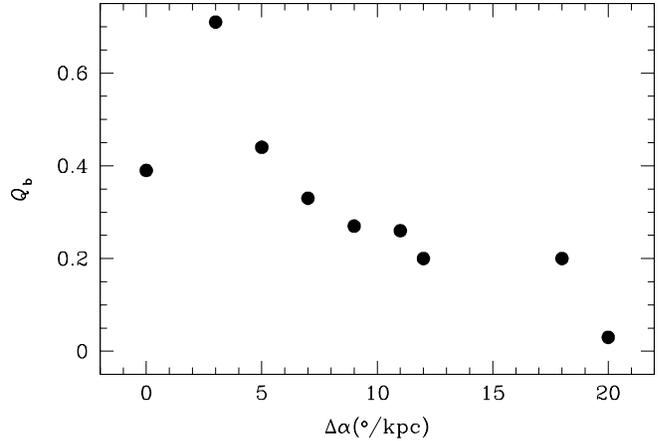},height=6cm}
\caption{As Fig.~10, but now with bar strength $Q_{\rm b}$.Typical
uncertainty  in $Q_{\rm b}$ is 0.1.}
\end{figure}

In this scheme, the lowest values of $\Delta \alpha$ correspond to the
straightest class of dust lanes (e.g., NGC~6951), whereas the highest
values obviously correspond to the most curved lanes (e.g.,
NGC~5248). The measurements were made along the largest possible stretch
of dust lane, which in all cases runs from just outside the CNR to very
near the end of the bar. Since dust lanes have rather constant curvature
over this entire range, use of a smaller part of the lanes to derive
$\Delta \alpha$ would lead to the same values, but with larger
uncertainties simply because the linear range involved would be smaller.
We estimate typical errors in $\Delta \alpha$ to be of the order of
3\deg/kpc, where uncertainties in the distance to the galaxy, and in the
definition of the curvature of the dust lane from the images are the
main contributing factors. The resulting values are tabulated in Table~5
and plotted against deprojected bar ellipticity in Fig.~10, and against
$Q_{\rm b}$ in Fig.~11.

The theoretically expected trend of more curvature of the dust lanes in
weaker bars is borne out clearly in the run of $Q_{\rm b}$ vs. dust lane
curvature, but much less clearly so, if at all, in the run of bar
ellipticity. This is in fact the first time that a correlation between
bar strength and dust lane curvature has been {\it observed} for a
sample (albeit admittedly small) of bars. We thus confirm
observationally the prediction from theory and modelling (e.g.,
Athanassoula 1992b). The fact that the correlation shows up much less
clearly when bar ellipticity is considered does not contradict this
confirmation, given the poor correlation between bar ellipticity and
$Q_{\rm b}$ shown by Block et al. (2001). We looked for, but did not
find, correlations between the curvature of the dust lanes and
parameters like host galaxy type or inclination angle.

\section{Summary}

This paper is the second of a pair presenting results of our study of
how the general structure of barred galaxies is related to, and possibly
determines, the presence and properties of dynamical features such as
rings and spirals in their CNRs.  We present optical broad and
narrow-band imaging of our sample galaxies in $B, I$ and H$\alpha$ in
order to map the structural components across the disks of these barred
galaxies. The 12 galaxies in our sample are all nearby spiral galaxies
known to host star-forming circumnuclear ring-like structure. We present
graphically our optical images and the radial profiles we derived from
these by ellipse fitting. From these data, we infer geometric and
dynamical parameters such as bar lengths and ellipticities, and location
of SF and general dust structures, and combine these with parameters
obtained from the literature.

Our main results can be summarised as follows: 

\begin{itemize}

\item For the first time using bar strengths ($Q_{\rm b}$ values from
the literature) rather than ellipticity, and using NIR imaging of the
bars, we checked the previously found correlation between bar strength
and length, i.e., longer bars are also stronger. Our results are
confusing because bar length as such correlates with bar strength, but
bar length relative to, e.g., the size of the galaxy does not. Careful
study of larger samples is needed.

\item The FIR index log($S_{25}$/$S_{100}$), a SF indicator, is used 
to verify whether the strength of the bar is connected to the global SF
activity. We find no relation with the bar strength $Q_{\rm b}$ nor with
the length of the bar.

\item We measure the relative H$\alpha$ flux from the disk, bar
and CNR components in the galaxies. We find that the more centrally
concentrated the \ha\ emission is (i.e., the higher the fraction of
total SF from the CNR), the more SF takes place in absolute terms in the
galaxy, as estimated from the FIR flux ratios. This is possibly due to
the IRAS index being more sensitive to CNR SF, which is concentrated in
a dust-rich environment. Alternatively, the CNR SF activity as measured
from \ha\ is related to the overall SF activity, as measured in the FIR,
of the disk. We find no correlation between the relative H$\alpha$ flux
from the disk, the bar, and the CNR with bar strength, nor with bar
length.

\item We explore relations between the strength of the bar and the
geometric parameters of the nuclear rings, namely their size relative to
the galaxy disk, intrinsic inclination, and PA offset between the major
axes of the nuclear ring and the bar which hosts it. We find that
stronger bars tend to host smaller rings, but that the bar strength is
unrelated to both the PA offset and the intrinsic ring ellipticity. The
fact that smaller rings occur in stronger bars can be understood from
bar orbit theory, which predicts that stronger bars limit the region
where $x2$ orbits can exist, and thus in turn the size of the nuclear
ring. As known from numerical modelling, both PA offset and intrinsic
ring ellipticity depend primarily on the mass inflow rate into the
nuclear ring, which in turn is determined only in part by the strength
of the bar, but more directly on the availability of fuel, and on
temporal variations therein.

\item We present a new, reproducible, measure of the curvature of dust
lanes in bars, a parameter which was theoretically linked about a
decade ago to the strength of the bar. This link is through gas
shocking under the influence of the bar potential, at the location
where pairs of symmetric dust lanes are seen along the bars in optical
images. We indeed confirm that the stronger the bar, the straighter
its dust lanes. This is in fact the first observational confirmation
of this theoretical result, based on the use of both a sample
(admittedly still rather small) of galaxies, and of quantitative
measures of both dust lane curvature and bar strength.

\end{itemize}

From these results, we can confirm that indeed the CNRs of barred
galaxies are intimately linked to their host galaxies, and that their SF
and morphological properties are determined to a significant degree by
the properties of, mainly, the large bar. The small size of our sample
precludes a detailed exploration of the correlations that we have
found. Studies using larger samples of nuclear ring galaxies are needed
to more fully test the predictions and results from theory and
modelling, and to explore further relations between the parameters
governing the CNR, the bar, and the disk.

{\it Acknowledgements}

We thank Milagros Ru\'\i z and Sharon Stedman for help during the
observing runs and for discussions, and Reynier Peletier for his help in
the preparation of Figure~1. We thank Isaac Shlosman for illuminating
discussion on bars and rings, and David Axon for discussions about
subtracting the H$\alpha$ continuum. We thank Ron Buta for calculating
$Q_{\rm b}$ values for NGC~1530 and NGC~6951. An anonymous referee is
acknowledged for comments which greatly helped to improve the
presentation of our results. This research has made use of the NASA/IPAC
Extragalactic Database (NED) which is operated by the Jet Propulsion
Laboratory, California Institute of Technology, under contract with the
National Aeronautics and Space Administration. The Jacobus Kapteyn and
Isaac Newton Telescopes are operated on the island of La Palma by the
ING in the Spanish Observatorio del Roque de los Muchachos of the
Instituto de Astrof\'\i sica de Canarias. Data were partly retrieved
from the ING archive.

\setcounter{figure}{0}

\begin{figure*}
\caption{
Images and photometric profiles for the 12  galaxies in our
sample. The lower left panel is a contour plus greyscale plot of the
galaxy in the $I$-band.  Contour levels are indicated in the captions to 
the individual galaxy figures. The coordinates are J2000.0, obtained from the
NASA/IPAC Extragalactic Database (NED). The panels in the middle show
a greyscale representation of the $B-I$ colour index (below) and
H$\alpha$ images. A close-up of the centre of the H$\alpha$
image is presented on the upper left, showing exactly the same region
as that shown for the NIR imaging in fig.~1 in Paper~I, to facilitate
intercomparison. Darker tones indicate redder colours in the $B-I$
image, more flux in H$\alpha$.  The right panels show radial profiles
of, from bottom  to top, $B$, $I$ and H$\alpha$ surface brightness,
$B-I$ colour, PA and ellipticity. The $B$ and $I$-band surface
brightness and $B-I$ colour have been photometrically calibrated using
aperture photometry from the literature, while the H$\alpha$ surface
brightness is uncalibrated and given in instrumental units of
log(counts). All profiles are plotted as a function of major axis
radius. The PA has been measured in degrees from N through
E. Small filled dots in the PA and ellipticity profiles are the NIR
values reproduced from fig.~1 of Paper~I. Uncertainties due to the
fitting procedure have not been indicated in the figure 
because they are smaller than
the symbol size in all cases, except for the colour index profile at
large radii, where uncertainties  can amount to $0.1-0.3$\,mag for the
outer few points plotted.
{\bf a.} NGC~1300. Contours from $I$=21 to 20\,mag\,arcsec$^{-2}$, with
intervals of 0.25 mag\,arcsec$^{-2}$, and from 19.5 to 17 with intervals
of 0.5; greyscale levels in $I$ from 21.5 to 17 in steps of 0.25.}

\end{figure*}

\setcounter{figure}{0}

\begin{figure*}
\caption{{\bf b.} NGC 1530. Contours and grey levels from $I$=21.5 to
17\,mag\,arcsec$^{-2}$ in steps of 0.5 mag\,arcsec$^{-2}$.}
\end{figure*}

\setcounter{figure}{0}

\begin{figure*}
\caption{{\bf c.} NGC 2903. NIR PA and ellipticity were not fitted in
Paper~I and have not been plotted. Contours and grey levels from $I$=20.0 to
16.25\,mag\,arcsec$^{-2}$ in steps of 0.25 mag\,arcsec$^{-2}$.}
\end{figure*}

\setcounter{figure}{0}

\begin{figure*}
\caption{{\bf d.} NGC 3351. Contours and grey levels from $I$=20.5 to
16\,mag\,arcsec$^{-2}$ in steps of 0.25 mag\,arcsec$^{-2}$. Differences
between the PSFs of the $B$ and $I$ images cause some small artifacts in 
the $B-I$ image.}
\end{figure*}

\setcounter{figure}{0}

\begin{figure*}
\caption{{\bf e.} NGC 3504. Contours and grey levels from $I$=21.5 to
15.5\,mag\,arcsec$^{-2}$ in steps of 0.5 mag\,arcsec$^{-2}$.}
\end{figure*}

\setcounter{figure}{0}

\begin{figure*}
\caption{{\bf f.} NGC 3516. Contours and grey levels from $I$=21.5 to
15.5\,mag\,arcsec$^{-2}$ in steps of 0.5 mag\,arcsec$^{-2}$.}
\end{figure*}

\setcounter{figure}{0}

\begin{figure*}
\caption{{\bf g.} NGC 3982. Values for $\mu_{B,I}$ and $B-I$ are
averages as determined from the other sample galaxies, the data for this 
galaxy have themselves not been calibrated. Contours  
from $I$=17.2 to
16.4\,mag\,arcsec$^{-2}$ in steps of 0.2 mag\,arcsec$^{-2}$, grey levels 
starting at $I$=19.0\,mag\,arcsec$^{-2}$ but otherwise the same.}
\end{figure*}

\setcounter{figure}{0}

\begin{figure*}
\caption{{\bf h.} NGC 4303. Contours and grey levels from $I$=20.75 to
16.0\,mag\,arcsec$^{-2}$ in steps of 0.25 mag\,arcsec$^{-2}$.}
\end{figure*}

\setcounter{figure}{0}

\begin{figure*}
\caption{{\bf i.} NGC 4314. Values for $\mu_{I}$ and $B-I$ are
averages as determined from the other sample galaxies, the data for this 
galaxy have themselves not been calibrated. Contours and grey levels 
from $I$=20.0 to
15.5\,mag\,arcsec$^{-2}$ in steps of 0.25 mag\,arcsec$^{-2}$.}
\end{figure*}

\setcounter{figure}{0}

\begin{figure*}
\caption{{\bf j.} NGC 4321. The nuclear region in the $I$-band image
is saturated, and no data are available in that region in the $I$ and 
$B-I$ images and the $I$ and $B-I$ profiles. Contours and grey levels 
from $I$=22.0 to
17.5\,mag\,arcsec$^{-2}$ in steps of 0.5 mag\,arcsec$^{-2}$.}
\end{figure*}

\setcounter{figure}{0}

\begin{figure*}
\caption{{\bf k.} NGC 5248. The nuclear region in the $I$-band image
is saturated, and no data are available in that region in the $I$ and 
$B-I$ images and the $I$ and $B-I$ profiles. Contours and grey levels 
from $I$=21.0 to
16.5\,mag\,arcsec$^{-2}$ in steps of 0.5 mag\,arcsec$^{-2}$}
\end{figure*}

\setcounter{figure}{0}

\begin{figure*}
\caption{{\bf l.} NGC 6951. The nuclear region in the $I$-band image is
saturated, and no data are available in that region in the $I$ and $B-I$
images and the $I$ and $B-I$ profiles. Values for $\mu_{I}$ and $B-I$
are averages as determined from the other sample galaxies, the data for
this galaxy have themselves not been calibrated. Contours and grey
levels from $I$=19.25 to
16.5\,mag\,arcsec$^{-2}$ in steps of 0.25 mag\,arcsec$^{-2}$.} 
\end{figure*}

\setcounter{figure}{1}

\begin{figure*}
\centerline{
}
\caption{{\bf a.} H$\alpha$ contours over a greyscale $J-K$ colour map
of NGC~1300. Contour levels are linearly increasing in instrumental
counts. Darker tones in the colour index image indicate redder
colours. {\bf b.} As Fig.~2a, now for NGC~1530.}
\end{figure*}

\setcounter{figure}{1}

\begin{figure*}
\centerline{
}
\caption{{\bf c.} As Fig.~2a now for NGC~2903. {\bf d.} H$\alpha$
contours (as in Fig.~2a) overlaid on a greyscale representation $K$-band
image of NGC~3351.}
\end{figure*}

\setcounter{figure}{1}

\begin{figure*}
\centerline{
}
\caption{{\bf e.} As Fig.~2a now for NGC~3504. {\bf f.} As Fig.~2a,
 now for NGC~3516.}
\end{figure*}

\setcounter{figure}{1}

\begin{figure*}
\centerline{
}
\caption{{\bf g.} As Fig.~2a now for NGC~3982. {\bf h.} As Fig.~2a,
 now for NGC~4303.}
\end{figure*}

\setcounter{figure}{1}

\begin{figure*}
\centerline{
}
\caption{{\bf i.} As Fig.~2a now for NGC~4314. {\bf j.} As Fig.~2a,
 now for NGC~4321.}
\end{figure*}

\setcounter{figure}{1}

\begin{figure*}
\centerline{
}
\caption{{\bf k.} As Fig.~2a now for NGC~5248. {\bf l.} As Fig.~2a,
 now for NGC~6951.}
\end{figure*}

\appendix

\section{Notes on individual galaxies}

\subsection{NGC 1300} 

This barred galaxy is a prototype grand-design SBb spiral. It has a very
prominent and smooth bar and two spiral arms that start abruptly at the
end of the bar (Fig.~1a).  In our $B-I$ image, we can see two straight,
offset dust lanes rather parallel to the major axis of the bar. These
dust lanes are smooth when compared to the arms, which corresponds to a
lack of distinct emitting regions. It is also apparent from our colour
map that, near the ends of the bar, the bar dust lanes join the dust
lanes along the inner edge of the arms.  SF is not obvious in the bar,
nor in the nucleus. The CNR is a site of considerable SF, concentrated
in a small area. There is a concentration of H{\sc ii} regions where the
arms join the bar, and mainly so on the western side. The western arm
also contains several very luminous H{\sc ii} regions, while the eastern
arm shows some less luminous enhancements.  The H$\alpha$ profile shows
a modest peak of luminosity at radii corresponding to those of the CNR,
and a gradual decrease in the region of the bar. The radial profiles
reflect and confirm the distribution of SF in this galaxy.

The nuclear ring in NGC~1300 (Fig.~1a, 2a) is rather small angularly, and
our \ha\ image does not give much information. It is clear, however,
that the massive SF occurs in a number of discrete clumps (see also
Pogge 1989), which are not reproduced in the NIR colour index images.
Whereas the nuclear ring is incomplete in \ha, it appears complete in
the NIR, although in Paper~I we showed that the NIR ring is in fact a
pair of tightly wound spirals. The {\it HST} image (fig.~1 of Paper~I)
shows structure in the nuclear ring, but no hot spots corresponding to
the \ha\ emission peaks.

\subsection{NGC 1530} 

NGC~1530 (Fig.~1b) has a large and moderately elliptical bar and two
wide open spiral arms originating from its ends.  Its most prominent
feature, however, is a bright mini-spiral at a radius of 10$\arcsec$
from the nucleus, and extended roughly perpendicularly to the bar. This
feature can be seen best in the \ha\ and $B-I$ images.  From CO and
H{\sc i} observations (e.g., Downes et al. 1996; Regan et al. 1996) it
is known that a large amount of gas is present in the centre. Despite
this, the galaxy has a relatively modest SF rate near the nucleus, which
can be explained by the trapping of gas in $x_{2}$ orbits near an inner
ILR, inhibiting further infall. The gas will thus not reach the critical
density that is needed to make the galaxy a true starburst (Downes et
al. 1996). In our H$\alpha$ image, we find relatively weak SF sites
along the bar, at the ends of the bar, and along the spiral arms. In the
CNR we find two distinct strongly emitting regions, from where the arms
of the mini-spiral seem to start. The H$\alpha$ profile shows a peak at
the centre, dips in the region of the bar, and a slight increase at
roughly the radius where the bar joins the disk spiral arms.  Our $B-I$
image shows two curved dust lanes emerging from the nucleus, which
continue along the bar.  Again, as in NGC~1300, it is apparent from our
colour map that, near the ends of the bar, the bar dust lanes join those
along the inner edge of the arms.

The two symmetrically placed clumps of massive SF (Fig.~1b, 2b) are not
obvious in the colour map, and seem to be hidden by the spiral
structure. They appear similar to K1 and K2 in M100 (Knapen et
al. 1995a), however, whereas in M100 K1 and K2 are most obvious in the
NIR, the peaks are not conspicuous at all in our NIR images of NGC~1530,
not even in the {\it HST} image (Paper~I). \ha\ emission is seen to
follow the northern miniature spiral arm, coinciding with the red spiral
arm fragment in the NIR colour index image.

\subsection{NGC 2903} 

NGC~2903 is a ``hot-spot'' galaxy with a circumnuclear starburst.  The
$B-I$ colour map reveals a beautiful and complex multi-armed dust
structure, but no clear pair of dust lanes in the bar (Fig.~1c). In the
H$\alpha$ images we see SF along the bar and in the arms, mostly in the
southern arm, and in the CNR. Very luminous H{\sc ii} regions delineate
the bar and arms in this galaxy. The H$\alpha$ profile drops smoothly,
outlining continuous and strong H{\sc ii} emission all the way from the
centre to the ends of the bar, and further out into the disk.  The
profiles show a lot of structure throughout the CNR and disk, caused by
the combined effects of dust extinction and SF activity. Blue dips in
the $B-I$ profile are accompanied by enhancements in the \ha\ and in
some cases the $B$ profiles. No NIR PA and ellipticity profiles are
given because these could not be fitted in Paper~I.

The \ha\ emitting regions do not all correspond in position and
intensity to the NIR hot spots (Fig.~2c). Our \ha\ image shows just
three of them, whereas our ground-based NIR images, and especially the
{\it HST} NIR image (Paper~I), show a plethora of individual emitting
regions. A detailed study of the SF properties of the CNR in this
galaxy, using {\it HST} archival \ha\ and Pa~$\alpha$ imaging and NIR
spectroscopy, in combination with stellar population synthesis
modelling, has recently been published by Alonso-Herrero, Ryder \&
Knapen (2001).

\subsection{NGC 3351} 

NGC~3351 shows a nuclear ring and a larger, inner, ring of H{\sc ii}
regions that encircles the stellar bar (Fig.~1d).  The $B-I$ image shows
two straight dust lanes emerging from an elliptical ring-like region
surrounding the nuclear ring.  The SF in this galaxy is significant and
concentrated in the CNR and the inner ring. From the H$\alpha$ profile
we can see a constant, high level in the nucleus, and a decrease along
the bar.  The profiles display a pronounced peak, coincident with the
location of the nuclear ring, at around 10 arcsec radius, and a
secondary peak corresponding to the inner ring. Both these peaks appear
blue in the $B-I$ colour plot reflecting the underlying SF.

In the close-up of the H$\alpha$ image of the CNR, the circumnuclear
ring is clearly visible, outlined by clumps of SF. For NGC~3351 we have
only a ground-based $K$-band image at our disposal, along with an {\it
HST} $H$-band image (Paper~I). The correspondence of the SF clumps of
its nuclear ring as seen in \ha\ with those seen in the NIR is obvious
from Fig.~2d, where the H$\alpha$ contours are shown overlaid on the
$K$-band image, and can be further traced in the {\it HST} image.

\subsection{NGC 3504} 

NGC 3504 is an inclined barred galaxy with a nuclear starburst. The
double nuclear peak seen in the NIR (Paper~I) does not show up in our
optical images due to the limited spatial resolution.  The $B-I$ colour
map shows two curved dust lanes in the bar region, but very little
organised dust in the disk region (Fig.~1e).  In our H$\alpha$ image, we
see an elliptical region around the nucleus, where SF activity is
intense. Outside the CNR, the most luminous H{\sc ii} regions are found
near the ends of the bar, but SF is also prominent in the rest of the
surrounding disk structure.  The radial H$\alpha$ profile shows a
decrease from the nucleus to the radius corresponding to the end of the
bar, where enhanced SF activity causes a local maximum in the
profile. The $B-I$ radial profile shows a blue peak at the same radius,
of about 50$\arcsec$. The values for the $B-I$ colour are between 1.5
and 2.  The central \ha\ emission in NGC~3504 (Fig.~2e) is elongated in
the direction of the double nucleus, as discussed in Paper~I.

\subsection{NGC 3516}

This galaxy hosts a Seyfert~1.5 nucleus, but shows little structure in
the optical.  The images displayed in Fig.~1f have a field of view
roughly half as big as the rest of the galaxies, due to the larger
distance and smaller angular size of this galaxy. The H$\alpha$ image
shows enhanced emission from the nuclear region, not necessarily from SF
but quite possibly from the AGN.  We do not detect any sign of SF in the
bar or disk of this galaxy.  The $B-I$ colour index image shows no
convincing structure in the disk, except possibly two redder regions
near the ends of the (main) bar. The nucleus is red, as seen in the
$B-I$ image and profile, but the white patches seen in the image
(Fig.~1f) may be artificial and induced by the nearby presence of the
strong nuclear peak. The $B-I$ profile decreases from the nucleus, and
becomes stable at 2.5 magnitudes.  The \ha\ emission in the CNR of
NGC~3516 (Fig.~2f) is featureless apart from a slight extension in the
E-W direction.

\subsection{NGC 3982}

NGC~3982 is one of two barred spirals with a Seyfert~2 nucleus in our
sample. The presence of a small bar, at a radius of around 10~arcsec, is
evident from the $I$-band image (Fig.~1g) and from the ellipticity and
PA profiles. Further out, the disk is dominated by a spiral pattern.
The $B-I$ colour index image displays a rich multi-armed spiral pattern
in dust and SF, but is rather smooth near the nucleus. This multi-armed
spiral pattern is clear also in the H$\alpha$ image, with several
luminous clumps of SF located mainly in the arm south of the nucleus,
and slightly less in the arm north of it. The nucleus of this galaxy,
while neutral in $B-I$, is a relatively strong H$\alpha$ emitter. We
cannot say how much of this emission is due to stars or the AGN, but the
latter is a likely contributor.  The H$\alpha$ profile displays a peak
in the nucleus, followed by a dip and a secondary peak corresponding in
radius to the zone where the SF sites in the spiral arms are
located. The profile drops rapidly after a radius of 20-25 arcsec. The
$B-I$ profile reflects the \ha\ peak due to the spiral arm SF as a blue
depression.  The calibration of our $B$ and $I$-band data is subject to
larger uncertainties than in other galaxies, because no aperture
photometry was available for this galaxy. We use average values for the
$B$ and $I$ surface brightness and $B-I$ colour, as derived from those
other sample galaxies for which calibration was available, and
concentrate on the changes in colours rather than on the exact
calibrated value.

The CNR of NGC~3982 (Fig.~2g), which may in fact be interpreted as the
emission from the relatively compact disk of this galaxy (the scale
length of the exponential disk is only 0.6~kpc - Table~1), shows many
individual sites of \ha\ and NIR emission. The relationship between
these two kinds of emission is not clear, however, due to the relatively
small signal to noise ratio in the areas of interest. Comparison with
the {\it HST} image as shown in Paper~I, however, indicates that the
same individual regions are responsible for
\ha\ and NIR emission.

\subsection{NGC 4303} 

The second Seyfert~2 host galaxy in our sample shows a well-defined bar
in the $I$-band image (Fig.~1h). The $B-I$ image shows two dust lanes in
the bar region, which join the inner edge of the spiral arms, and
continue as arm dust lanes. The radial profiles in \ha\ and $B$ show a
central peak in surface brightness, followed by a decrease to the radius
corresponding to the end of the bar where a secondary peak is located,
and a steeper decrease in the disk. The $B-I$ profile reflects these
zones as well.  The H$\alpha$ image shows SF activity along the arms and
at the ends of the bar, where the bar joins the arms. An especially
large concentration of H{\sc ii} regions is located near the northern
end of the bar, and a smaller number of H{\sc ii} regions, but of higher
luminosity, are found near the southern end. No SF is visible in the bar
itself, but the CNR (Fig.~2h) is the site of substantial H$\alpha$
emission, the morphology of which is however not fully resolved with our
imaging. The $B-I$ image reflects in blue colours where these main sites
of SF are. The fact that the CNR, so bright in \ha, is not as blue as
the regions near the ends of the bar may indicate substantial amounts of
dust in the central regions. Alternatively, a substantial fraction of
the CNR \ha\ emission is due to the Seyfert nucleus and not to
circumnuclear SF. We refer to Colina et al. (1997) and Colina \& Arribas
(1999) for more detailed imaging of the CNR of this galaxy.

\subsection{NGC 4314} 

NGC~4314 is a symmetrical SBa galaxy with a long stellar bar and two
faint outer spiral arms, not visible in our rather shallow $I$-band
image (Fig.~1i).  The H$\alpha$ image shows a number of H{\sc ii}
regions forming an incomplete bright nuclear ring, and substantial
emission from the (LINER) nucleus. There is no sign of SF activity in
the bar or disk of this galaxy. In fact, 94\% of the total H$\alpha$
emission from the galaxy originates in the central region (Pogge 1989).
In the $B-I$ colour map, the bar is rather featureless, but in the CNR
we see several well-defined blue knots corresponding to the sites of SF,
and a pair of short dust lanes which depart from the circumnuclear
ring. In the NIR colour index images in Paper~I, these dust lanes were
seen to form a smooth and continuous ring. The presumed continuation of
these dust lanes into the bar cannot be distinguished in the image.  The
morphology of the galaxy can clearly be recognised in the H$\alpha$
profile: a nuclear peak followed by a depression, a secondary peak at
the radius of the star-forming CNR, and a rapid fall-off outside the
CNR. The CNR can be recognised as a dip in the $B-I$ profile, as well as
in the $B$-band profile as a peak due to enhanced blue emission. We find
no evidence for a double bar in this galaxy, neither from our own NIR
(Paper~I) and optical data, nor from the HST NIR image shown in
paper~I. The ellipticity does go up slightly around 10~arcsec in radius,
but this rise is accompanied by large changes in PA; these changes are
most probably caused by the strongly emitting distinct regions in the
CNR. We thus do not confirm the suggestion by Benedict et al. (1993) and
Friedli et al. (1996) of a secondary bar with a semi-major axis length
of around 4~arcsec.  As in the case of NGC~3982, no aperture photometry
in the $I$-band have been published. We thus took an average calibration
number from the rest of our sample and applied this to the $I$ and $B-I$
profiles.

The \ha\ image of the CNR of NGC~4314 (Fig.~2i) displays four main
discrete clumps of massive SF in an incomplete ring. The ring is
visible, and appears more complete, in the $J-K$ colour map, which shows
no evidence for discrete SF sites. The \ha\ hot spots are located very
near or on the red ring-like region in the colour index map, which
points towards an origin of the NIR red colours there in massive SF. The
{\it HST} image (Paper~I) shows that at high resolution the NIR ``ring''
is resolved into many individual hot spots, located along a pair of very
tightly wound spiral arms.

\subsection{NGC 4321} 

NGC 4321 (M100) is a late-type barred galaxy with well-defined spiral
arms and a spectacular central region which only reveals some of its
hidden features through the use of the NIR observations. Knapen et
al. (1995a,b) discussed in detail the existence and origin of a stellar
bar with a large-scale and a small-scale component, a two-armed
star-forming spiral in the CNR, and a pair of leading arms inside the
nuclear ring-like region.  We have used the same optical images as used
by Knapen \& Beckman (1996) and refer the reader to that paper for a
detailed discussion. In summary, the $B-I$ colour index map shows a pair
of curved dust lanes in the bar continuing into the disk spiral arms.  A
string of H{\sc ii} regions accompanies the dust lanes in the bar. There
is abundant H$\alpha$ emission from the CNR and the disk of this
galaxy. Knapen (1998) catalogued almost 2000 individual H{\sc ii}
regions from the image shown here and presents a statistical analysis of
their properties, but also discusses the H$\alpha$ morphology.  Knapen
\& Beckman (1996) presented surface brightness profiles for, among other
tracers, $B, I, B-I$ and H$\alpha$. Although the profiles in the present
paper were determined using a different ellipse fitting algorithm, we
confirm the main characteristics of the surface brightness and colour
profiles. The ellipticity and PA profiles confirm the two parts of the
bar (Fig.~1j).

As discussed in detail in the literature (e.g., Knapen et al. 1995a,b,
2000b), images of the CNR of NGC~4321 (Fig.~2j) show good coincidence
between the features highlighted by the $J-K$ colour map and the
H$\alpha$ contours. We can see clearly the SF clumps along the arms of
the miniature spiral structure (including the characteristic K1 and K2
points -- red loci accompanied by massive SF toward the SE and NW of the
nucleus).

\subsection{NGC 5248} 

Laine et al. (1999) used some of the optical data presented here, in
combination with our NIR data from Paper~I and adaptive optics NIR
imaging, to describe the spiral structure in this galaxy at various
scales, including at scales of tens of pc, where it manifests itself in
the form of a nuclear grand-design spiral. We refer the reader to Laine
et al. for a description of the general disk morphology of NGC~5248, as
well as of the CNR. Further detailed studies of the CNR and host galaxy
can be found in Laine et al. (2001) and Jogee et al. (2002a,b).  Our
$B-I$ colour index map (Fig.~1k) shows a pair of curved dust lanes
emerging from the nuclear region, more prominent southwest than
northeast of the nucleus. Massive SF, as measured by H$\alpha$ emission,
is very strong in the CNR and in the (non-active) nucleus, and is
located mainly along the spiral arms in the bar (see Jogee et
al. 2002a).  The H$\alpha$ profile outlines the central region with its
peak in intensity, followed by a second maximum that corresponds to the
region of the arms where the SF is found. In the region of the second
maximum a small bump in the $B-I$ colour profile is also visible. The
ellipticity and PA profiles are remarkably flat.

NGC~5248 (Fig.~2k) also exhibits massive SF sites mainly situated in the
nuclear ring, seen in H$\alpha$ and in the {\it HST} $H$-band image
(Paper~I). The massive SF activity seems not to extend to the spiral
arms and is confined to the nuclear ring and the region immediately
surrounding it.

\subsection{NGC 6951} 

Our images of this galaxy (Fig.~1l) suffer from rather poor spatial
resolution, but the strong bar and main spiral arms in the disk can
easily be identified. The $B-I$ image shows a pair of curved dust lanes
which emerges from the CNR and runs roughly parallel to the spiral arms.
The enhanced SF activity in the CNR is obvious from the blue colour in
the $B-I$ map and from the H$\alpha$ image, but the resolution in the
images is too poor to show much detail (compare, e.g., to the $H$-band
HST image shown in Paper~I). The H$\alpha$ image shows SF activity
across the disk, but none is observed along the bar. There is also some
H$\alpha$ emission from the (LINER/Seyfert) nucleus.

We encountered two problems in the process of the ellipse fitting,
resulting in higher than normal uncertainties: the original images had
saturated centres, and no aperture photometry for the $I$-band was
found. We thus used average values for $\mu_I$ and $B-I$ as labelled in
the appropriate profiles in Fig.~1l. We see no evidence for an inner bar
from our combined optical and NIR profile fits, contrary to the
suggestion of Wozniak et al. (1995) as based on optical imaging, a
suggestion which, however, could not be unambiguously confirmed by the
same team from their NIR imaging (Friedli et al. 1996).

NGC~6951 (Fig.~2l), like NGC~5248, shows miniature spiral
structure in the {\it HST} image, but this spiral is not resolved in our 
\ha\ image and the clumps of massive SF seem to be confined to the
ring. As in, e.g.,  NGC~4314, the regions of massive SF as traced by
\ha\ emission coincide in radius with the red ``ring'' as traced by the
$J-K$ image.

\bsp


\end{document}